\documentclass[notitlepage,a4,10.5pt]{article}

\usepackage{graphicx}

\usepackage{amsmath}%
\usepackage{amsfonts}%
\usepackage{amssymb}%
\usepackage{mathrsfs}
\usepackage{amsthm}
\usepackage{float}

\usepackage{authblk}

\usepackage[left=2.5cm,right=2.5cm,top=3cm,bottom=3cm]{geometry} 

\usepackage{xcolor}
\usepackage{stmaryrd}

\newcommand{\mc}{\mathcal}
\newcommand{\cp}{\times}

\newcommand{\bol}{\boldsymbol}

\newcommand{\abs}[1]{\left\lvert{#1}\right\rvert}
\newcommand{\w}{\wedge}
\newcommand{\lr}[1]{\left({#1}\right)}
\newcommand{\lrs}[1]{\left[{#1}\right]}
\newcommand{\lrc}[1]{\left\{{#1}\right\}}

\newcommand{\mf}{\mathfrak}
\newcommand{\p}{\partial}

\newcommand{\ti}[1]{\textit{#1}}
\newcommand{\tb}[1]{\textbf{#1}}

\begin{document}

\title{A Collision Operator for Describing Dissipation\\ in Noncanonical Phase Space}
\author[1]{Naoki Sato} 
\author[2]{Philip J. Morrison}
\affil[1]{National Institute for Fusion Science, \protect\\ 322-6 Oroshi-cho Toki-city, Gifu 509-5292, Japan \protect\\ Email: sato.naoki@nifs.ac.jp}
\affil[2]{Department of Physics and Institute for Fusion Studies, \protect\\ 
University of Texas, Austin, Texas
78712, USA
 \protect \\ Email: morrison@physics.utexas.edu}
\date{\today}
\setcounter{Maxaffil}{0}
\renewcommand\Affilfont{\itshape\small}

    \maketitle
    \begin{abstract}
    The phase space of a noncanonical Hamiltonian system is partially inaccessible due to 
    dynamical constraints (Casimir invariants) arising from the kernel of the Poisson tensor. 
    When an ensemble of noncanonical Hamiltonian systems is allowed to interact, dissipative processes eventually break the phase space constraints, resulting in a thermodynamic equilibrium described by a Maxwell-Boltzmann distribution. 
    However, the time scale required to reach Maxwell-Boltzmann statistics is often much longer than the time scale over which a given system achieves a state of thermal equilibrium. 
    Examples include diffusion in rigid mechanical systems, as well as  
    collisionless relaxation in magnetized plasmas and stellar systems, where the interval between binary Coulomb or gravitational collisions can be longer than the time scale over which stable structures are self-organized.   
    Here, we focus on self-organizing phenomena over spacetime scales such that particle interactions respect the noncanonical Hamiltonian structure, but yet act to create a state of thermodynamic equilibrium. 
    We derive a collision operator for general noncanonical Hamiltonian systems,  
    applicable to fast, localized interactions.  
    This collision operator depends on the interaction exchanged by colliding particles and on the Poisson tensor encoding the noncanonical phase space structure, is consistent with entropy growth and conservation of particle number and energy, preserves 
    the interior Casimir invariants, reduces to the Landau collision operator in the limit of grazing binary Coulomb collisions in canonical phase space, and exhibits a metriplectic structure. We further show how thermodynamic equilibria depart from Maxwell-Boltzmann statistics due to the noncanonical phase space structure, and how self-organization and collisionless relaxation in magnetized plasmas and stellar systems can be described through the derived collision operator.  
    \end{abstract}

\section{Introduction}


This paper is concerned with the following problem in kinetic theory. Suppose that the distribution function $f\lr{\bol{z},t}$ of a statistical ensemble with phase space coordinate $\bol{z}$ evolves in time $t$ according to
\begin{equation}
\frac{\p f}{\p t}=-\lrc{f,\mf{H}}_{\ast}+\lr{\frac{\p f}{\p t}}_{\rm coll},
\end{equation}
where the Hamiltonian functional $\mf{H}$ represents the total energy of the system,  $\lrc{\cdot,\cdot}_{\ast}$ is a noncanonical Poisson bracket \cite{Morrison98,Morrison82} for the field theory, and $\lr{\p f/\p t}_{\rm coll}$ denotes a collision term accounting for the interaction between particles during scattering events. 
What is the expression of the collition term $\lr{\p f/\p t}_{\rm coll}$ for binary collisions that occur over short time scales and small spatial scales compared to the ideal dynamics described by the noncanonical Poisson bracket? 
Here, we show that the collision term is given by
\begin{equation}
\lr{\frac{\p f}{\p t}}_{\rm coll}=\mc{C}\lr{f,f}=\frac{\p}{\p\bol{z}}\cdot\lrs{f\mc{J}\cdot\int f'\Pi\cdot\lr{\mc{J}'\cdot\frac{\p\log f'}{\p\bol{z}'}-\mc{J}\cdot\frac{\p\log f}{\p\bol{z}}}d\bol{z}'},\label{eq2}
\end{equation}
where `$\cdot$' means contraction, $\Pi$ is a symmetric covariant $2$-tensor (the interaction tensor), whose form depends on the type of binary interactions, and $\mc{J}$ denotes the contravariant Poisson $2$-tensor (bivector) associated with the noncanonical Poisson bracket. 
In this notation, $f'=f\lr{\bol{z}',t}$ and $\mc{J}'=\mc{J}\lr{\bol{z}'}$. 

In order to explain the physical meaning of this result, 
we start by recalling the role played by noncanonical Hamiltonian mechanics in the description of dynamical systems. 
Physical phenomena are characterized by spacetime scales. 
These scales impart a dynamical hierarchy to the degrees of freedom of a system. 
Depending on the phenomenon of interest, part of these degrees of freedom may be redundant, and therefore  inessential to the description of the evolution of the system. For example, the dynamics of a rigid body is conveniently described by the three components of angular momentum, while momenta and positions of the microscopic constituents of the rigid body can be removed from the mathematical formulation of the equations of motion. 
Noncanonical Hamiltonian mechanics represents the mathematical structure arising  
from such process of reduction. Notice that this mechanical framework is not the result of a change of coordinates (in the rigid body example, the phase space is three-dimensional, and therefore it cannot be spanned by paired canonical variables), but the intrinsic structure of reduced phase spaces.  
It is therefore legitimate to ask what is the statistical behavior of an ensemble of noncanonical Hamiltonian systems, such as an ensemble of interacting rigid bodies. 
Under suitable working assumption on the nature of the interaction driving dissipation, this question can be reformulated as the mathematical problem of determining the collision operator that appears within the Boltzmann transport equation when, however, the underlying phase space structure is noncanonical.
This is the theoretical issue addressed in this work. 

The physical motivation behind this study stems from the observation that certain physical systems, such as magnetized plasmas in the laboratory \cite{YosPRL} or in  astrophysical environments (e.g. accretion disks or planetary radiations belts \cite{Kawazura19}) or systems of gravitating bodies (e.g. stars in the galactic disk \cite{Lynden}), often self-organize into equilibria over time scales that are $(i)$  insufficient to break the phase space obstructions represented by Casimir invariants or $(ii)$ sensibly shorter than the typical time interval (collision time) between two scattering events. In fact, one speaks of self-organization \cite{Yos2014} and collisionless relxation \cite{Chavanis22,Kadomstev}. 
Here, self-organization refers to the creation of ordered structures in thermodynamically closed systems subject to contraints, while collisionless relaxation describes an entropy increasing process in which binary collisions between particles populating a statistical ensemble do not contribute to the formation of equilibria. In this context, a particle is  
an elementary contituent of a statistical ensemble, 
such as a charged particle, 
a rigid body, a star, and so on. 
It should also be emphasized that self-organization and collisionless relaxation are not mutually exclusive, nor mutually inclusive.  
These self-organizing/relaxation phenomena raise two fundamental questions: $(i)$ since many of these systems can be effectively modeled as thermodynamically closed, why are non-trivial structures formed during relaxation and how are they consistent with entropy growth? $(ii)$ if binary collisions are absent, what is the relaxation mechanism enabling the creation of a self-organized equilibrium state?

The results reported in this work suggest that the answer to these questions lies in the noncanonical Hamiltonian structure of the phase space associated with microscopic dynamics. Indeed, the constrained nature or total absence of binary collisions
ensures that the noncanonical structure of the equations of motion is preserved during the relaxation process, and the resulting thermal equilibria turn out to depend explicitly on the geometry of noncanonical phase space. This is, we argue, the mechanism by which self-organized structures can be created while maximizing entropy. 
It is only over longer time scales over which the usual binary collisions become dominant that the noncanonical structure is destroyed, the contraints removed, and the usual Maxwell-Boltzmann statistics is recovered. 

Furthermore, the collisionless relaxation mechanism can be understood in terms of collective fluctuations associated with the ensemble averaged interaction potential energy felt by the 
particles, in the same way mean electromagnetic and gravitational fields determine the evolution of the distribution function in the Vlasov-Maxwell and Vlasov-Poisson models \cite{Morrison82,MorrisonMV,Marsden82}. 
This relaxation process can be modeled as a binary collision process between localized `clusters' of a large number of particles because the cluster collision frequency is a linear function of the particles contained in a cluster, and therefore becomes sensibly larger than the usual collision frequency. Mathematically, the process can be described in terms of a collision operator analogous to the Landau collision operator \cite{Landau1936,Kampen,Lenard} for grazing Coulomb collisions in canonical phase space. 
In this setting, the particle distribution function $f$ is replaced by the distribution function of particle clusters, $\tilde{f}$, 
and the kinetic equation satisfied by $\tilde{f}$
effectively generalizes the Landau kinetic equation to noncanonical phase spaces. We may identity $\tilde{f}$ and $f$ with the notions of coarse-grained distribution function and fine-grained distribution function that often occur in statistical mechanics.  
We also remark that the notion of particle clusters adopted in this paper to describe collisionless relaxation is conceptually analogous to the particle clumps occurring in the context of phase space density granulation theory of plasmas (see e.g. \cite{Dupree72}), 
and shares similarities with the quasiparticles arising in the dressed test particle model \cite{Rostoker},
but it is not limited to the Coulomb interaction. 
Furthermore, we stress again that the applicability of the present theory is not limited to collisionless relaxation, because it deals with the broader issue of the description of dissipation and collisions in general noncanonical Hamiltonian systems. 

There are two other aspects that we must discuss before constructing the collision operator in noncanonical phase space: the context of this work with respect to the wider framework of statistical mechanics and kinetic theory, and the methodology used to tackle the theoretical problem. First, we remark that the present work is not concerned with the identification of new types of statistics, such as those associated with non-additive entropies \cite{Chavanis04,Presse} or those arising from indistinguishability of particles and exclusion principles as in the Lynden theory \cite{Lynden} or Bose-Einstein and Fermi-Dirac statistics, but rather with the problem of how 
noncanonical phase space may alter the usual understanding of ergodicity and differential entropy.  
More precisely, due to the presence of constraints (the Casimir invariants of the Poisson bracket \cite{Sudar,Littlejohn}, which represent constants of motion that are independent of the Hamiltonian (energy) of the system) and the nontrivial nature of the invariant measure (preserved phase space measure) in a noncanonical Hamiltonian system, 
the accessible regions of the phase space are restricted, and the phase space metric is distorted with respect to the usual  
configuration space metric. 
These obstructions restrict the ergodicity \cite{Moore} of noncanonical Hamiltonian systems to submanifolds corresponding to level sets of the Casimir invariants, and affect 
the definition of Shannon's differential entropy $S=-\int f\log f\,d\Omega$ \cite{Shannon},  which is a non-covariant functional of the distribution function \cite{Jaynes,Sato2016}. 
We remark here that the key role played by Casimir invariants in shaping 
ordered structures in self-organizing phenomena has been pointed out in several contexts, including magnetized plasmas \cite{Yos2014}, collisionless systems with long-range interactions \cite{Lynden,Chavanis06}, and two-dimensional ideal fluid flows \cite{Jung,Chavanis96}. 
In a field theory
arising from a microscopic dynamical system 
Casimir invariants can be classified into interior (or induced)  Casimir invariants and outer Casimir invariants. 
Interior Casimir invariants are inherited from the kernel of the Poisson tensor associated with microscopic dynamics 
by the Poisson bracket of the corresponding field theory. On the other hand, an outer Casimir invariant is a property of the Poisson bracket of the field theory, such as the total particle number in the Vlasov-Poisson model. 
We will see that in the kinetic theory developed
in the present paper a critical role is played by the interior Casimir invariants, which are responsible for deviation from Maxwell-Boltzmann statistics.  

With regard to the proper notion of entropy in noncanonical Hamiltonian systems, we note that the argument of the logarithm appearing in the definition of $S$, which must be a pure probability, depends on the choice of the volume element $d\Omega$ with respect to which the probability density $f$ is defined. Hence, the 
functional $S$ represents a complete information measure only if the volume element $d\Omega$ is invariant when transported by the dynamical flow. 

The starting point of the construction developed in this work is the BBGKY  hierarchy of equations \cite{Kampen} (see  \cite{Marsden} for the Hamiltonian structure) for the $i$-point distribution function $f_i\lr{\bol{z}_1,...,\bol{z}_i,t}$ of a system consisting of $N$ particles with phase space positions $\bol{z}_i$, $i=1,...,N$, of which we report the first equation
\begin{equation}
\frac{\p f_1}{\p t}\lr{\bol{z},t}+\lrc{f_1,H_1+\Phi_1\lrs{f_1}}\lr{\bol{z},t}=\int\lrc{f_1\lr{\bol{z},t}f_1\lr{\bol{z}',t}-f_{2}\lr{\bol{z},\bol{z}',t},V\lr{\bol{z},\bol{z}'}}\,d\bol{z}',\label{BBGKY1}
\end{equation}
where $H_1\lr{\bol{z}}$ is the energy of an isolated particle, 
$\Phi_1\lrs{f_1}=\int Vf'_1\,d\bol{z}'$ the ensemble averaged interaction energy, $V\lr{\bol{z},\bol{z}'}$ the potential energy of binary interactions (which is assumed to be symmetric in its arguments), and $\lrc{\cdot,\cdot}$ denotes the one-particle Poisson bracket. The relationship between the Poisson bracket of the field theory  $\lrc{\cdot,\cdot}_{\ast}$ and the one-particle Poisson bracket $\lrc{\cdot,\cdot}$ is discussed in section 7.
When binary collisions between particles are negligible (in the sense that their occurrence is rare compared to the characteristic time scale under consideration) the term on the right-hand side of equation \eqref{BBGKY1} can be neglected, because the probability $f_2\lr{\bol{z},\bol{z}',t}d\bol{z}d\bol{z}'$ of finding two particles in the phase space volume elements centered at $\bol{z}$ and $\bol{z}'$ can be approximated in terms of the product  $f_1\lr{\bol{z},t}f_1'\lr{\bol{z}',t}d\bol{z}d\bol{z}'$ due to the 
small correlation of the events. In this case, particle interactions are mediated solely by the ensemble averaged potential energy $\Phi_1$,   
and one recovers the Vlasov model (on this point, see also \cite{Kadomstev,Ewart} where the construction of effective collision operators for the Vlasov equation in canonical phase space is discussed). 
Collision operators can then be introduced by a suitable closure assumption  
relating the $2$-point distribution function to the $1$-point distribution function in the term on the right-hand side of equation \eqref{BBGKY1}. 

However, as we will see in sections 2 and 3, 
the construction of the collision operator becomes a nontrivial theoretical challenge   
when the Poisson bracket $\lrc{\cdot,\cdot}$ is noncanonical because its properties will not only depend on the nature of the particle interaction, but also on the geometry of the phase space.   
Furthermore, the notions of particle cluster and cluster collisions are required to model collisionless relaxation, since the term on the right-hand side of \eqref{BBGKY1} is negligible in this setting. 
Indeed, suppose that we construct the BBGKY hierarchy for a system of $N_{\rm cl}=N/N'$ clusters of particles with $i$-point distribution function $\tilde{f}_i\lr{\bol{z}_1,...,\bol{z}_i,t}$. If the number of particles $N'$ contained in each cluster is high enough, the 
correlation $\tilde{f}_1\lr{\bol{z},t}\tilde{f}_1\lr{\bol{z}',t}-\tilde{f}_2\lr{\bol{z},\bol{z}',t}$ is no longer negligible because 
the larger phase space volume occupied by each cluster increases the likelihood of binary cluster collisions.
Then, a kinetic model of the collision process can be obtained by 
closing the hierarchy equations in terms of a collision operator for the distribution $\tilde{f}_1$, which is one of the tasks we undertake in this study. 

With regard to the derivation of the collision operator, we emphasize that in the context of constrained or collisionless relaxation the noncanonical Hamiltonian structure encapsulated in the Poisson bracket survives during the relaxation process due to the constrained or negligible nature of usual collisions, and therefore appears explicitly in the expression of the collision operator 
through the Poisson tensor $\mc{J}$ as shown in equation \eqref{eq2}. We also note that $\mc{J}$ always appears multiplied by $\mc{J}'$ in the collision operator, a behavior that is reminiscent but different from the case of so-called double brackets \cite{Vallis,Bloch,Fl,Furukawa} used as a practical method to compute equilibria in noncanonical Hamiltonian systems. Instead, it will be shown that the kinetic equation satisfied by the distribution function 
possesses a metriplectic structure \cite{Morrison1984,Morrison1986}, i.e. an algebraic structure 
associated with dissipative systems that is consistent with the first and second laws of thermodynamics. 
This algebraic structure has been found in several physical systems spanning different phenomena, such as dissipative magnetohydrodynamics \cite{Materassi,Coquinot} or the Fokker-Planck equation \cite{Sato2020}. Due to the generality of the dynamical systems examined, this work suggests that the metriplectic bracket 
characterizes the phase space of dissipative systems in the same way the Poisson structure characterizes the phase space of ideal systems. 

We remark that although collisionless relaxation represents one of the physical motivations, the present theory applies to noncanonical Hamiltonian systems in general. On the other hand, collisionless or weakly collisional systems cannot be necessarily described within the present framework, nor they always exhibit stable structures (see e.g. \cite{Sche} and \cite{Kawazura20}). Indeed, the range of validity of the developed theory depends on the nature of the collision process, which must be fast and spatially confined with respect to the spacetime scales associated with the underlying noncanonical Hamiltonian structure. It is also worth mentioning that the problem examined here is conceptually different from 
the derivation of the collision operator in gyrokinetic theory \cite{Hirv,Burby}, where 
the expression of the cyclotron angle averaged Landau collision operator is sought in terms of the reduced gyrokinetic phase space variables. 
Nevertheless, the generality of the approach developed in this work may provide useful insight into the problem. 
 

The present paper is organized as follows. 
In section 2, we describe the mechanical setting of collisions in noncanonical phase space. 
In section 3, we derive the collision operator in noncanonical phase space by expanding a collision integral in powers of the scattering displacements experienced by the colliding particles. The procedure is conceptually analogous to the one used to obtain the Landau collision operator from the Boltzmann collision integral.
In section 4, we first show that the derived collision operator satisfies conservation of total particle number, energy (unlike double brackets), and Casimir invariants. Then, we prove an H-thereom, from which we obtain the form of thermodynamic equilibria. 
In section 5, the application of the theory to self-organization 
in constrained mechanical systems and collisionless magnetized plasmas and steallar systems is discussed. 
In section 6, we show that the derived collision operator reduces to the Landau collision operator in the limit of grazing Coulomb collisions in canonical phase space. 
In section 7, we exhibit the metriplectic bracket of the 
derived kinetic equation.
Concluding remarks are given in section 8. 



\section{Binary interactions in noncanonical phase space}
We begin by considering a noncanonical Hamiltonian system described by the equations of motion
\begin{equation}
\dot{\bol{z}}=\mc{J}\p_{\bol{z}}H=\mc{J}^{ij}\frac{\p H}{\p z^j}\p_i.\label{EoM}
\end{equation}
In this notation $\bol{z}=\lr{z^1,...,z^n}$ 
are coordinates 
spanning an $n$-dimensional domain $\Omega\subseteq\mathbb{R}^n$, 
$\mc{J}\in\bigwedge^2 T\Omega$ a Poisson $2$-tensor with components $\mc{J}^{ij}=-\mc{J}^{ji}$, $i,j=1,...,n$,  $H=H\lr{\bol{z}}$ the Hamiltonian function, 
$\p_i$ the tangent vector in the $z^i$ direction,  and $t\in[0,\infty)$ the time variable such that  
$\dot{\bol{z}}=d\bol{z}/dt$. 
Furthermore, 
we have simplified the notation by omitting the contraction symbol in  $\mc{J}\cdot\p_{\bol{z}}H$. This convention will often be used in the rest of the paper. 

It is convenient to regard equation \eqref{EoM} as the equations of motion of a `particle' with energy $H$, so that the notion of collisions among particles is analogous to the familiar one encountered in the settings of the Boltzmann or Landau collision operators. It should be emphasized however that the kind of collisions that will be considered in this study may be essentially different from elastic scatterings between pairs of real particles. Indeed, the colliding particles may represent more general interacting  systems, such as two clusters of real charged particles interacting with each other via the electromagnetic force. Furthermore, interactions will not be restricted to scatterings in velocity space.   

For the purpose of the present paper we will assume that collisions occur between two ensembles consisting of the same type of particles, e.g. electrons colliding with electrons. This hypothesis greatly simplifies the algebra of the expansion of the collision integral, but is not essential for the following theory to hold. 
Now consider two particles with coordinates $\bol{z}_1$ and $\bol{z}_2$ and energies $H_1=H\lr{\bol{z}_1}$ and $H_2=H\lr{\bol{z}_2}$ respectively. 
Let $V=V\lr{\bol{z}_1,\bol{z}_2}$
denote the energy of the interaction between them. 
The energies of the interacting particles therefore read as
\begin{equation}
\mc{H}_1=H_1+V,~~~~\mc{H}_2=H_2+V, 
\end{equation}
while the respective equations of motion accounting for the interaction are
\begin{equation}
\dot{\bol{z}}_1=\mc{J}_1\p_{\bol{z}_1}\mc{H}_1,~~~~\dot{\bol{z}}_2=\mc{J}_2\p_{\bol{z}_2}\mc{H}_2.\label{EoM2}
\end{equation}
Here, $\mc{J}_1=\mc{J}\lr{\bol{z}_1}$ and $\mc{J}_2=\mc{J}\lr{\bol{z}_2}$. 
Notice that in these equations 
the two particles share the same phase space structure through the same Poisson tensor $\mc{J}$. 
However, different Poisson tensors arise if the colliding species are different. Next, observe that the energies $\mc{H}_1$ and $\mc{H}_2$ obey 
\begin{equation}
\frac{d\mc{H}_1}{dt}=\dot{\bol{z}}_1\cdot\p_{\bol{z}_1}\mc{H}_1+\dot{\bol{z}}_2\cdot\p_{\bol{z}_2}\mc{H}_1=\mc{J}_2\p_{\bol{z}_2}H_2\cdot\p_{\bol{z}_2}V,~~~~\frac{d\mc{H}_2}{dt}=\dot{\bol{z}}_1\cdot\p_{\bol{z}_1}\mc{H}_2+\dot{\bol{z}}_2\cdot\p_{\bol{z}_2}\mc{H}_2=\mc{J}_1\p_{\bol{z}_1}H_1\cdot\p_{\bol{z}_1}V.
\end{equation}
On the other hand, the total energy of the two-particle system
\begin{equation}
\mc{H}_{12}=H_1+V+H_2,
\end{equation}
satisfies
\begin{equation}
\frac{d\mc{H}_{12}}{dt}=\frac{d\mc{H}_1}{dt}+\frac{dH_2}{dt}=\mc{J}_2\p_{\bol{z}_2}H_2\cdot\p_{\bol{z}_2}V+\mc{J}_2\p_{\bol{z}_2}V\cdot\p_{\bol{z}_2}H_2=0,\label{Htot} 
\end{equation}
where we used the antisymmetry of $\mc{J}$. Note that $V$ does not need to satisfy $\p_{\bol{z}_1}V=-\p_{\bol{z}_2}V$ for \eqref{Htot} to hold. 

Further assumptions are needed on the nature of the interaction described by $V$ for the evaluation of the collision operator that will be carried out in the next section. Let 
\begin{equation}
L\sim \frac{\mc{J}^{ij}}{\frac{\p\mc{J}^{ij}}{\p z^k}}\sim\frac{H}{\frac{\p H}{\p z^j}},~~~~T\sim \frac{L}{\mc{J}^{ij}\frac{\p H}{\p z^j}}, 
\end{equation}
denote the spatial and time scales associated with the Poisson tensor $\mc{J}$ and the Hamiltonian $H$. 
Similarly, let
\begin{equation}
\ell_c\sim \frac{V}{\frac{\p V}{\p z^j}},~~~~\tau_c\sim\frac{\ell_c}{\mc{J}^{ij}\frac{\p V}{\p z^j}},
\end{equation}
denote the spatial and time scales of the collision process. 
In the following, we assume that collisions involve small spacetime scales according to
\begin{equation}
\frac{\ell_c}{L}<<1,~~~~\frac{\tau_c}{T}<<1.\label{scales}
\end{equation}
We remark that in the setting of the Landau collision operator for Coulomb collisions within a plasma, 
the scale length $\ell_c$ is essentially the Debye length $\lambda_D$, while 
the time scale $\tau_c$ can be estimated through the ratio $\lambda_D/\sqrt{k_BT/m}$, with $T$ and $m$ the plasma temperature and charged particle mass respectively.

In addition, we restrict our attention to interaction energies $V$ that result in elastic scatterings, i.e. collision events that preserve the total energy of the involved particles prior to the interaction,
\begin{equation}
E=H_1+\Phi_1+H_2+\Phi_2,\label{E}
\end{equation}
where we have introduced 
the ensemble averaged interaction energies  
\begin{equation}
\Phi_1=\Phi_1\lr{\bol{z}_1,t}=\int_{\Omega} f_2\lr{\bol{z}_2,t}V\lr{\bol{z}_1,\bol{z}_2}d\bol{z}_2,~~~~\Phi_2=\Phi_2\lr{\bol{z}_2,t}=\int_{\Omega} f_1\lr{\bol{z}_1,t}V\lr{\bol{z}_1,\bol{z}_2}d\bol{z}_1.
\end{equation}
Here, $f_1\lr{\bol{z}_1,t}$ and $f_2\lr{\bol{z}_2,t}$ denote the probability density functions (distribution functions) of the colliding ensembles defined with respect to the phase space measures $d\bol{z}_1$ and $d\bol{z}_2$ respectively.  
Of course, if collisions occur between particles belonging to the same ensemble with probability density function $f\lr{\bol{z},t}$, one has $f_1\lr{\bol{z}_1,t}=f\lr{\bol{z}_1,t}$ and $f_2\lr{\bol{z}_2,t}=f\lr{\bol{z}_2,t}$. 
We also remark that $\Phi_1$ represents the interaction energy of the first colliding particle with all particles belonging to the other ensemble, excluding the second colliding particle (it is assumed that,  due to the large number of particles populating each statistical ensemble, the subtraction of a particle leaves the corresponding distribution function essentially unchanged). 
A similar interpretation applies to $\Phi_2$.
Differentiating \eqref{E} with respect to time, one obtains  
\begin{equation}
\begin{split}
\frac{dE}{dt}=&\p_{\bol{z}_1}{\lr{H_1+\Phi_1}}\cdot\dot{\bol{z}}_1+\p_{\bol{z}_2}\lr{H_2+\Phi_2}\cdot\dot{\bol{z}}_2\\
=&\p_{\bol{z}_1}\lr{H_1+\Phi_1}\cdot\mc{J}_{1}\p_{\bol{z}_1}\lr{E+V}+
\p_{\bol{z}_2}\lr{H_2+\Phi_2}\cdot\mc{J}_{2}\p_{\bol{z}_2}\lr{E+V}.\label{dEdt}
\end{split}
\end{equation}
Here, we used the fact that the equations of motion can be written as $\dot{\bol{z}}_1=\mc{J}_{1}\p_{\bol{z}_1}\lr{E+V}$ and $\dot{\bol{z}}_2=\mc{J}_{2}\p_{\bol{z}_2}\lr{E+V}$. 
The rate of change \eqref{dEdt} therefore vanishes 
provided that the following elastic scattering condition is satisfied:
\begin{equation}
\p_{\bol{z}_1}{\lr{H_1+\Phi_1}}\cdot\mc{J}_1\p_{\bol{z}_1}V+\p_{\bol{z}_2}\lr{H_2+\Phi_2}\cdot\mc{J}_2\p_{\bol{z}_2}V=0.\label{dEdt2}
\end{equation}
We further assume the following symmetry condition for the potential energy $V$,  
\begin{equation}
\frac{\p V}{\p\bol{z}_1}=-\frac{\p V}{\p\bol{z}_2},\label{sym}
\end{equation}
so that the elastic scattering condition \eqref{dEdt2} reduces to
\begin{equation}
\left[\mc{J}_1\p_{\bol{z}_1}{\lr{H_1+\Phi_1}}-\mc{J}_2\p_{\bol{z}_2}\lr{H_2+\Phi_2}\right]\cdot\p_{\bol{z}_1}V=0.\label{dEdt3}
\end{equation}
Here, it should be noted that $\Phi_1$ and $\Phi_2$ depend  on the distribution functions. However, since $V$ usually appears as a function of spatial coordinates, and the Poisson tensor links spatial coordinates to momentum coordinates, terms involving $\Phi_1$ and $\Phi_2$ are expected to vanish in equation \eqref{dEdt3}. 

The condition \eqref{sym} has the following  interpretation:  
in a physical process the forces exchanged by the particles 
must balance each other. These forces correspond to the gradient of the potential $V$ in configuration space: 
the interaction force acting on a particle at $\bol{z}_1$ 
is $\bol{F}_1=-\p_{\bol{q}_1}V=-\frac{\p V}{\p z_1^i}\frac{\p z^i_1}{\p\bol{q}_1}$.  
Newton's third law therefore reads
\begin{equation}
\bol{F}_1+\bol{F}_2=-\frac{\p V}{\p{z}_1^i}
\frac{\p {z}_1^i}{\p\bol{q}_1}-\frac{\p V}{\p{z}_2^i}
\frac{\p {z}_2^i}{\p\bol{q}_2}=\bol{0}.
\end{equation}
However, on the small scale $\ell_c$ of the interaction $\p_{\bol{q}_1} z_1^i\approx\p_{\bol{q}_2} z_2^i$, leading to \eqref{sym}.

As an example, 
in the case of Coulomb scatterings among charged particles, one has $\bol{z}=\lr{\bol{v},\bol{q}}$ where $\bol{v}=\dot{\bol{q}}$ is the particle velocity and $\bol{q}$ the particle position, $H_1=m\bol{v}_1^2/2$, $\Phi_1=\Phi_1\lr{\bol{q}_1}$, $H_2=m\bol{v}_2^2/2$, $\Phi_2=\Phi_2\lr{\bol{q}_2}$, $V=V\lr{\abs{\bol{q}_1-\bol{q}_2}}$, $\p_{\bol{q}_1}V=-\p_{\bol{q}_2}V$, with $\lr{m\bol{v}_1,\bol{q}_1}$ and $\lr{m\bol{v}_2,\bol{q}_2}$ canonical variables
and $m$ the particle mass. 
It follows that the elastic scattering condition \eqref{dEdt3} becomes
\begin{equation}
\lr{{\bol{v}_1}-{\bol{v}_2}}\cdot\p_{\bol{q}_1}V=
0,\label{eq20}
\end{equation}
which expresses the conservation of the modulus of the relative velocity $\abs{{\bol{v}}_1-{\bol{v}}_2}$ during a binary Coulomb scattering. 
In summary, the Hamiltonian $H$, the Poisson tensor $\mc{J}$,  and the interaction energy $V$
must be chosen so that the conditions \eqref{sym} and \eqref{dEdt3} are satisfied for the theory developed in the present paper to hold. 
Observe that \eqref{sym} implies $V=V\lr{\bol{z}_1-\bol{z}_2}$, as one can verify through the change of variables $\lr{\bol{z}_1,\bol{z}_2}\rightarrow\lr{\bol{z}_1-\bol{z}_2,\bol{z}_1+\bol{z}_2}/2$.

Returning to the general case, we observe that, due to the 
 spatially localized nature of the interaction (the energies $H_1+\Phi_1$ and $H_2+\Phi_2$ and the Poisson tensors $\mc{J}_1$ and $\mc{J}_2$ are effectively constant over the spatial scale of the collision $\ell_c$ thanks to the first ordering hypothesis in \eqref{scales}), the shortness of the collision (the second ordering hypothesis in \eqref{scales}), and the symmetry of the interaction energy \eqref{sym}, 
the changes in the phase space positions $\bol{z}_1$ and $\bol{z}_2$ due to a collision event can be approximated as 
\begin{equation}
\delta\bol{z}_1\approx \tau_c\mc{J}_1\p_{\bol{z}_1}V\approx -\tau_c\mc{J}_2\p_{\bol{z}_2}V=-\delta\bol{z}_2.\label{dx0}
\end{equation}
Hence, the total displacement vanishes, i.e.
\begin{equation}
\delta\bol{z}_1+\delta\bol{z}_2\approx\bol{0}.\label{dx}
\end{equation}
We will see that the property \eqref{dx} enables a construction analogous to that of the Landau collision operator by simplifying the expansion of the collision integral in powers of $\delta\bol{z}_1$ and $\delta\bol{z}_2$. 

The last ingredient needed to evaluate the collision operator is to demand that the coordinate system $\bol{z}=\lr{z^1,...,z^n}$ defines an invariant measure, that is
\begin{equation}
\mf{L}_{\dot{\bol{z}}}d\bol{z}=
\frac{\p}{\p z^i}\lr{\mc{J}^{ij}\frac{\p H}{\p z^j}}d\bol{z}=0~~~~\forall H,\label{IM}
\end{equation}
where $\mf{L}$ denotes the Lie-derivative and $d\bol{z}=dz^1...dz^n$ the phase space measure. 
Since $\mc{J}$ is antisymmetric and $H$ is arbitrary, equation \eqref{IM} is equivalent to demanding that the column of $\mc{J}$ are divergence-free, 
\begin{equation}
\frac{\p\mc{J}^{ij}}{\p z^j}=0,~~~~i=1,...,n.\label{IM2}
\end{equation}
An antisymmetric matrix $\mc{J}$ (not necessarily a Poisson tensor) satisfying \eqref{IM2} is called measure preserving. 
Equation \eqref{IM} deserves further clarification. 
First, we recall that in a canonical Hamiltonian system with $2m$ canonical coordinates $\lr{\bol{p},\bol{q}}=\lr{p^1,...,p^m,q^1,...,q^m}$
and canonical Poisson tensor given by 
\begin{equation}
\mc{J}_c=\begin{bmatrix}\bol{0}_m&-\bol{I}_m\\\bol{I}_m&\bol{0}_m\end{bmatrix},
\end{equation} 
where $\bol{0}_m$ and $\bol{I}_m$ are the $m$-dimensional
null and identity matrices, 
equation \eqref{IM} is identically satisfied for any choice of the Hamiltonian function $H$ because $\mc{J}_c$ is constant and antisymmetric. 
Next, observe that for a general noncanonical Hamiltonian system with Poisson tensor $\mc{J}$, the existence of an invariant measure $d\bol{z}$ satisfying \eqref{IM} is guaranteed locally by the Lie-Darboux theorem \cite{Darboux1,Darboux2}. More precisely, 
for a sufficiently regular $\mc{J}$ of dimension $n$, one can find a
small neighborhood $U\subseteq\Omega$ 
and a local system of coordinates $\lr{p^1,...,p^m,q^1,...,q^m,C^1,...,C^s}$, with $2m=n-s$ the rank of $\mc{J}$ in $U$, such that in these coordinates the matrix representation of $\mc{J}$ is given by 
\begin{equation}
\mc{J}=\begin{bmatrix}\bol{0}_m&-\bol{I}_m&\bol{0}_{ ms}\\\bol{I}_m&\bol{0}_m&\bol{0}_{ms}\\\bol{0}_{ sm}&\bol{0}_{sm}&\bol{0}_{s}\end{bmatrix},
\end{equation}
where $\bol{0}_{ms}$ is the null matrix with $m$ rows and $s$ columns. 
It readily follows that in such case an invariant measure satisfying \eqref{IM} in the neighborhood $U$ can be chosen as
\begin{equation}
d\bol{z}=dp^1...dp^mdq^1...dq^mdC^1...dC^s.
\end{equation}
It is worth emphasizing that the local coordinates $\lr{\bol{p},\bol{q}}$ play the role of the usual canonical pairs, while the coordinates $\bol{C}=\lr{C^1,...,C^s}$  correspond to 
$s$ constants of motion called Casimir invariants that hold regardless of the specific form of $H$. Indeed, they belong to the kernel of the Poisson tensor, so that 
\begin{equation}
\dot{C}^a=\frac{\p C^{a}}{\p z^i}\mc{J}^{ij}\frac{\p H}{\p z^j}=0~~~~\forall H,~~~~a=1,...,s.\label{Cas}
\end{equation}

In the following we shall assume that the coordinate system $\bol{z}$ is an invariant measure in the sense of equation \eqref{IM} in the whole volume $\Omega$. In practice, this means that the Poisson tensor $\mc{J}$ satisfies equation \eqref{IM2}. 
The existence of such Hamiltonian independent invariant measure is essential for the formulation of statistical mechanics because it implies that collisions, which are modeled by the interaction energy $V$, do not alter the preserved phase space volume $d\bol{z}$ (to see this, recall that the equations of motion accounting for the interaction are given by \eqref{EoM2}). We will see that this fact is crucial for the definition of differential entropy
\begin{equation}
S\left[f\right]=-\int_{\Omega}f\log f\,d\bol{z},\label{S}
\end{equation}
where $f\lr{\bol{z},t}$ denotes the probability density function with respect to the measure $d\bol{z}$, 
because the functional $S$ is not covariant (it depends on the chosen coordinate system).
In particular, performing a change of coordinates $\bol{z}\rightarrow\bol{y}$ such that $d\bol{z}={ J}\,d\bol{y}$, with ${J}$ the Jacobian determinant of the transformation, shows that
\begin{equation}
S\left[f\right]=-\int_{\Omega}g\log\lr{\frac{g}{J}}\,d\bol{y}=S\left[g\right]+\int_{\Omega}f\log J\,d\bol{z}\neq S\left[g\right],
\end{equation}
where $g\lr{\bol{y}}$ denotes the probability density function with respect to the measure $d\bol{y}$, i.e. $fd\bol{z}=gd\bol{y}$.


\section{Collision operator in noncanonical phase space}
We are now ready to construct the collision operator for the noncanonical Hamiltonian system described in the previous section. Let $f$ denote the probability density function (distribution function) of a particle species defined with respect to the invariant measure $d\bol{z}$. 
In the absence of interactions, the conservation of probability $\lr{\p_t+\mf{L}_{\dot{\bol{z}}}}fd\bol{z}=0$, with $\dot{\bol{z}}$ given by \eqref{EoM}, implies that the rate of change in $f$ obeys
\begin{equation}
\frac{\p f}{\p t}=-\frac{\p}{\p z^i}\lr{\mc{J}^{ij}\frac{\p H}{\p z^j}f}=-\frac{\p f}{\p z^i}\mc{J}^{ij}\frac{\p H}{\p z^j}=-\left\{f,H\right\},
\end{equation}
where we used equation \eqref{IM2} and introduced the Poisson bracket $\left\{\cdot,\cdot\right\}$ acting on pairs of functions $f,g$ according to
\begin{equation}
\left\{f,g\right\}=\frac{\p f}{\p z^i}\mc{J}^{ij}\frac{\p g}{\p z^j}.\label{PB} 
\end{equation}
Now suppose that an ensemble of particles with distribution $f_1\lr{\bol{z}_1,t}$ is allowed to interact with a second ensemble with distribution $f_2\lr{\bol{z}_2,t}$. The respective evolution equations read
\begin{subequations}
\begin{align}
\frac{\p f_1}{\p t}&=-\left\{f_1,{H}_1+\Phi_1\right\}_1+\mc{C}\lr{f_1,f_2},\\
\frac{\p f_2}{\p t}&=-\left\{f_2,{H}_2+\Phi_2\right\}_2+\mc{C}\lr{f_2,f_1}.
\end{align}
\end{subequations}
Here, $\left\{f,g\right\}_\alpha=\frac{\p f}{\p z^i}\mc{J}^{ij}_\alpha \frac{\p g}{\p z^j}$, $\alpha=1,2$,  
while $\mc{C}\lr{f_1,f_2}$ is the collision operator measuring the rate of change in $f_1$ due to collisions with the ensemble with distribution $f_2$.   
Our task is thus to evaluate $\mc{C}$ when the interaction driving collisions is given by the interaction energy $V$ introduced above. 
We emphasize that the ensemble averaged interaction energies $\Phi_1$ and $\Phi_2$ and the collision operators $\mc{C}\lr{f_1,f_2}$ and $\mc{C}\lr{f_2,f_1}$ represent distinct contributions to the rate of change in the distributions functions $f_1$ and $f_2$. Indeed, the former terms reflect the average modification of the particles energies caused by the presence of interactions, while the collision operators arise from the statistical correlation between $f_1$ and $f_2$, i.e. the non-vanishing of the difference $f_1\lr{\bol{z}_1,t}f_2\lr{\bol{z}_2,t}-f_{12}\lr{\bol{z}_1,\bol{z}_2,t}$, where $f_{12}$ denotes the distribution function of the combined ensemble (recall equation \eqref{BBGKY1} and see reference \cite{Marsden,Kampen} on this point). 
Neglecting such correlation is the working hypothesis leading to the Vlasov-Maxwell model \cite{Morrison82,MorrisonMV}. 

As usual, we assume that binary collisions are the largest contribution to $\mc{C}$. Nevertheless, we stress again that such binary scatterings do not need to take place between two real particles, but they may represent a binary interaction between more general systems, such as two clusters of charged particles. 
Assuming that the colliding particles are originally located at $\bol{z}_1$ and $\bol{z}_2$ respectively, we denote with 
$\bol{z}_1'$ and $\bol{z}_2'$ the respective phase space positions after a collision event. 
The rate at which particles of the ensemble $f_1$ leave $\bol{z}_1$ due to collisions with particles of the second ensemble $f_2$ to reach some new position $\bol{z}_1'$ can be expressed as
\begin{equation}
-\int\mc{V}\lr{\bol{z}_1,\bol{z}_2;\bol{z}_1',\bol{z}_2'}f_1\lr{\bol{z}_1,t}f_2\lr{\bol{z}_2,t}\,d\bol{z}_1'd\bol{z}_2'd\bol{z}_2,
\end{equation}
where the range of integration has been omitted to simplify the notation. Such convention will also be adopted in the rest of the paper.  
The quantity $\mc{V}d\bol{z}_1'd\bol{z}_2'd\bol{z}_2$ is the \ti{scattering volume per unit time}, i.e. the scattering phase space volume scooped by a particle at $\bol{z}_1$ in the unit time interval. We shall refer to the quantity $\mc{V}$, which has dimensions of $\left[d\bol{z}^{-2}t^{-1}\right]$, as the scattering volume density per unit time.
For the Boltzmann collision operator, one has $\bol{z}=\lr{\bol{v},\bol{q}}$ and also 
\begin{equation}
\mc{V}=m^{-6}\sigma\abs{{\bol{v}}_1-{\bol{v}}_2}\delta\lr{\bol{q}_1-\bol{q}_2}\delta\lr{\bol{q}_1'-\bol{q}_2'}\delta\lr{\bol{q}_1-\bol{q}_1'},\label{svc} 
\end{equation}
where $\sigma\lr{{{\bol{v}}}_1,{\bol{v}_2};{\bol{v}}'_1,{\bol{v}}'_2}$ is the scattering cross-section and 
${\bol{v}}_1$, ${\bol{v}}_2$, ${\bol{v}}'_1$, and ${\bol{v}}'_2$ are the spatial velocities of the scattered particles before and after the collision.  
In a similar fashion, the rate at which particles originally at some position $\bol{z}'_1$ reach $\bol{z}_1$ via scattering with $f_2$ is
\begin{equation}
\int\mc{V}\lr{\bol{z}'_1,\bol{z}'_2;\bol{z}_1,\bol{z}_2}f_1\lr{\bol{z}'_1,t}f_2\lr{\bol{z}'_2,t}\,d\bol{z}_1'd\bol{z}_2'd\bol{z}_2.
\end{equation}
On the other hand, the reversibility of the scattering process requires that
\begin{equation}
\mc{V}\lr{\bol{z}_1,\bol{z}_2;\bol{z}'_1,\bol{z}'_2}=\mc{V}\lr{\bol{z}'_1,\bol{z}'_2;\bol{z}_1,\bol{z}_2}.\label{rev}
\end{equation}
We therefore conclude that
\begin{equation}
\mc{C}\lr{f_1,f_2}=\int\mc{V}\lr{\bol{z}_1,\bol{z}_2;\bol{z}'_1,\bol{z}'_2}\left[f_1\lr{\bol{z}'_1,t}f_2\lr{\bol{z}'_2,t}-f_1\lr{\bol{z}_1,t}f_2\lr{\bol{z}_2,t}\right]\,d\bol{z}_1'd\bol{z}_2'd\bol{z}_2.\label{C0}
\end{equation}
Note that $\mc{C}\lr{f_1,f_2}\lr{\bol{z}_1,t}=\mc{C}\lr{f_2,f_1}\lr{\bol{z}_1,t}$ and that collisions are non-local in the whole phase space (this is in contrast with the Boltzmann or Landau collision operators which are local in position space). 
Next, recall that, by hypothesis, collisions occur over small spatial scales, i.e. the displacement
\begin{equation}
\delta\bol{z}=\delta\bol{z}_1=\bol{z}_1'-\bol{z}_1=-\delta\bol{z}_2=\bol{z}_2-\bol{z}_2',
\end{equation}
is small compared to the spatial scale $L$ 
characterizing Hamiltonian dynamics without interactions. 
Here, we used equation \eqref{dx}.
Hence, we may expand the distribution functions appearing in the integrand of the collision operator \eqref{C0} in powers of $\delta\bol{z}$ 
around $\bol{z}_1$ and $\bol{z}_2$. 
To this end, it is convenient to omit the arguments of the functions appearing in the collision integral to simplify the notation. For example, we shall write $f_1$ in place of $f_1\lr{\bol{z}_1,t}$. 
At second order in $\delta\bol{z}$ 
we obtain:
\begin{equation}
\begin{split}
\mc{C}\lr{f_1,f_2}=&
\int\mc{V}\left[\lr{f_1+\frac{\p f_1}{\p\bol{z}_1}\cdot\delta\bol{z}+\frac{1}{2}\delta\bol{z}\cdot\frac{\p^2f_1}{\p\bol{z}_1^2}\cdot\delta\bol{z}}\lr{f_2-\frac{\p f_2}{\p\bol{z}_2}\cdot\delta\bol{z}+\frac{1}{2}\delta\bol{z}\cdot\frac{\p^2f_2}{\p\bol{z}_2^2}\cdot\delta\bol{z}}-f_1f_2\right]\,d\bol{z}_1'd\bol{z}_2'd\bol{z}_2\\=
&\int\mc{V}\left[\lr{f_2\frac{\p f_1}{\p\bol{z}_1}-f_1\frac{\p f_2}{\p\bol{z}_2}}\cdot\delta\bol{z}
+\frac{1}{2}\delta\bol{z}\cdot\lr{f_1\frac{\p^2f_2}{\p\bol{z}_2^2}+f_2\frac{\p^2f_1}{\p\bol{z}_1^2}}\cdot\delta\bol{z}-{\frac{\p f_1}{\p\bol{z}_1}\cdot\delta\bol{z}}{\frac{\p f_2}{\p\bol{z}_2}\cdot\delta\bol{z}}\right]\,d\bol{z}_1'd\bol{z}_2'd\bol{z}_2.\label{C1}
\end{split}
\end{equation}
It is now useful to define the vector field
\begin{equation}
\bol{J}=f_2\frac{\p f_1}{\p\bol{z}_1}-f_1\frac{\p f_2}{\p \bol{z}_2}.
\end{equation}
Observing that
\begin{equation}
\frac{1}{2}\lr{\frac{\p}{\p\bol{z}_1}-\frac{\p}{\p\bol{z}_2}}\bol{J}=\frac{1}{2}f_2\frac{\p f_1}{\p \bol{z}_1^2}-\frac{1}{2}\frac{\p f_1}{\p\bol{z}_1}\frac{\p f_2}{\p\bol{z}_2}-\frac{1}{2}\frac{\p f_2}{\p\bol{z}_2}\frac{\p f_1}{\p\bol{z}_1}+\frac{1}{2}f_1\frac{\p^2f_2}{\p\bol{z}_2^2},
\end{equation}
equation \eqref{C1} then becomes
\begin{equation}
\begin{split}
\mc{C}\lr{f_1,f_2}=
\int\mc{V}\left\{\bol{J}\cdot\delta\bol{z}+\frac{1}{2}\delta\bol{z}\cdot\left[\lr{\frac{\p}{\p\bol{z}_1}-\frac{\p}{\p\bol{z}_2}}\bol{J}\right]\cdot\delta\bol{z}\right\}\,d\bol{z}_1'd\bol{z}_2'd\bol{z}_2.\label{C2}
\end{split}
\end{equation}
Next, notice that from \eqref{dx0} and \eqref{IM2} we have
\begin{equation}
\frac{\p}{\p\bol{z}_1}\cdot\delta\bol{z}_1=\tau_c\frac{\p}{\p\bol{z}_1}\cdot\lr{\mc{J}_1\cdot\frac{\p V}{\p\bol{z}_1}}=0,~~~~\frac{\p}{\p\bol{z}_2}\cdot\delta\bol{z}_2=\tau_c\frac{\p}{\p\bol{z}_2}\cdot\lr{\mc{J}_2\cdot\frac{\p V}{\p\bol{z}_2}}=0.
\end{equation}
Hence, after some algebraic manipulations, the collision operator can be expressed as
\begin{equation}
\begin{split}
\mc{C}\lr{f_1,f_2}=&\int\bol{J}\cdot\left[\mc{V}\delta\bol{z}-\frac{1}{2}\lr{\frac{\p}{\p\bol{z}_1}-\frac{\p}{\p\bol{z}_2}}\cdot\lr{\mc{V}\delta\bol{z}\delta\bol{z}}\right]\,d\bol{z}_1'd\bol{z}_2'd\bol{z}_2\\&+\frac{1}{2}\frac{\p}{\p\bol{z}_1}\cdot\int\mc{V}\delta\bol{z}\bol{J}\cdot\delta\bol{z}\,d\bol{z}_1'd\bol{z}_2'd\bol{z}_2-\frac{1}{2}\int\frac{\p}{\p\bol{z}_2}\cdot\lr{\mc{V}\delta\bol{z}\bol{J}\cdot\delta\bol{z}}\,d\bol{z}_1'd\bol{z}_2'd\bol{z}_2.\label{C3}
\end{split}
\end{equation}
Since interactions occur over small spatial scales $\ell_c<<L$, the displacement $\delta\bol{z}$ caused by the collision of a particle at  $\bol{z}_1$, sufficiently distant from the boundary, with a particle of the other ensemble, located on the boundary $\p\Omega$ of the region $\Omega$,  vanishes. It follows that the last term, which can be written as a boundary integral, is zero. Then,  
\begin{equation}
\begin{split}
\mc{C}\lr{f_1,f_2}=&\int\bol{J}\cdot\left[\mc{V}\delta\bol{z}-\frac{1}{2}\lr{\frac{\p}{\p\bol{z}_1}-\frac{\p}{\p\bol{z}_2}}\cdot\lr{\mc{V}\delta\bol{z}\delta\bol{z}}\right]\,d\bol{z}_1'd\bol{z}_2'd\bol{z}_2+\frac{1}{2}\frac{\p}{\p\bol{z}_1}\cdot\int\mc{V}\delta\bol{z}\bol{J}\cdot\delta\bol{z}\,d\bol{z}_1'd\bol{z}_2'd\bol{z}_2.\label{C4}
\end{split}
\end{equation}
Next, recall that the kind of elastic collisions considered here do not change the total number of particles. 
Hence, we must have
\begin{equation}
\int\frac{\p f_1}{\p t}\,d\bol{z}_1=\int\bol{J}\cdot\left[\mc{V}\delta\bol{z}-\frac{1}{2}\lr{\frac{\p}{\p\bol{z}_1}-\frac{\p}{\p\bol{z}_2}}\cdot\lr{\mc{V}\delta\bol{z}\delta\bol{z}}\right]\,d\bol{z}_1'd\bol{z}_2'd\bol{z}_1d\bol{z}_2=0.
\end{equation}
However, the only term depending on $f_1$ and $f_2$ within the integrand is $\bol{J}=\bol{J}\lr{\bol{z}_1,\bol{z}_2,t}$. Since there is no restriction on the particular shape of the distributions $f_1\lr{\bol{z}_1,t}$ and $f_2\lr{\bol{z}_2,t}$ at a given instant $t$, one therefore expects that
\begin{equation}
\int\left[\mc{V}\delta\bol{z}-\frac{1}{2}\lr{\frac{\p}{\p\bol{z}_1}-\frac{\p}{\p\bol{z}_2}}\cdot\lr{\mc{V}\delta\bol{z}\delta\bol{z}}\right]\,d\bol{z}_1'd\bol{z}_2'=\bol{0}.\label{VInt}
\end{equation}
Then, the collision operator \eqref{C4} reduces to
\begin{equation}
\mc{C}\lr{f_1,f_2}=\frac{1}{2}\frac{\p}{\p\bol{z}_1}\cdot\int\mc{V}\delta\bol{z}\bol{J}\cdot\delta\bol{z}\,d\bol{z}_1'd\bol{z}_2'd\bol{z}_2.\label{C5}
\end{equation}
 We now define the scattering frequency through the integral    
\begin{equation}
\Gamma_{12}=\Gamma\lr{\bol{z}_1,\bol{z}_2}=\int\mc{V}d\bol{z}_1'd\bol{z}_2'.
\end{equation}
Note that $\Gamma$ has dimensions of $\left[t^{-1}\right]$. 
We expect the scattering frequency for a particle $\bol{z}_1$ impinging on a target $\bol{z}_2$ to be the same for the conjugate process, so that 
\begin{equation}
\Gamma=\Gamma_{12}=\Gamma_{21}. 
\end{equation}
Using the expression \eqref{dx0} for the displacement $\delta\bol{z}$ 
one thus arrives at
\begin{equation}
\mc{C}\lr{f_1,f_2}=\frac{\tau_c^2}{2}\frac{\p}{\p\bol{z}_1}\cdot\left[f_1\mc{J}_1\cdot\int f_2\Gamma\frac{\p V}{\p\bol{z}_1}\frac{\p V}{\p\bol{z}_1}\cdot\lr{\mc{J}_2\cdot\frac{\p\log f_2}{\p\bol{z}_2}-\mc{J}_1\cdot\frac{\p\log f_1}{\p\bol{z}_1}}\,d\bol{z}_2\right].\label{C6}
\end{equation}
where we used the fact that on the spatial scale of the interaction $\mc{J}_1\approx\mc{J}_2$. 
It is useful to introduce the following symmetric covariant $2$-tensor
\begin{equation}
\Pi_{ij}=\frac{1}{2}\tau_c^2\Gamma\frac{\p V}{\p z_1^i}\frac{\p V}{\p z_1^j}=\frac{1}{2}\tau_c^2\Gamma\frac{\p V}{\p z_2^i}\frac{\p V}{\p z_2^j}.\label{Pi}
\end{equation}
This tensor, which we will call the interaction tensor, is a property of the interaction force driving  particle collisions. It can be regarded as a `degenerate metric tensor' (positive semi-definite tensor) whose kernel is spanned by vector fields belonging to the $n-1$ dimensional  tangent space $T\Sigma_V$ with $\Sigma_V=\lrc{\bol{z}\in\Omega:V\lr{\bol{z}}=c\in\mathbb{R}}$ the level sets of $V$. 

The collision operator \eqref{C6} can thus be equivalently expressed as  
\begin{equation}
\begin{split}
\mc{C}\lr{f_1,f_2}=&
\frac{\p}{\p\bol{z}_1}\cdot \left[f_1\mc{J}_1\cdot\int f_2\Pi\cdot\lr{\mc{J}_2\cdot\frac{\p\log f_2}{\p\bol{z}_2}-\mc{J}_1\cdot\frac{\p\log f_1}{\p\bol{z}_1}}d\bol{z}_2\right]
\\
=&\frac{\p}{\p z_1^i}\left[f_1\mc{J}_1^{ij}\int\Pi_{jk}\lr{\mc{J}^{km}_2\frac{\p\log f_2}{\p z_2^m}-\mc{J}^{km}_1\frac{\p\log f_1}{\p z_1^m}}d\bol{z}_2\right].\label{C7}
\end{split}
\end{equation}
The complete evolution equations for the distributions $f_1$ and $f_2$ therefore read as
\begin{subequations}
\begin{align}
\frac{\p f_1}{\p t}=&\frac{\p}{\p \bol{z}_1}\cdot\left\{f_1\mc{J}_1\cdot\left[-\frac{\p\lr{H_1+\Phi_1}}{\p \bol{z}_1}+\int f_2\Pi\cdot\lr{\mc{J}_2\cdot\frac{\p\log f_2}{\p\bol{z}_2}-\mc{J}_1\cdot\frac{\p\log f_1}{\p\bol{z}_1}}d\bol{z}_2\right]\right\},\label{f1t}\\
\frac{\p f_2}{\p t}=&
\frac{\p}{\p \bol{z}_2}\cdot\left\{f_2\mc{J}_2\cdot\left[-\frac{\p\lr{H_2+\Phi_2}}{\p \bol{z}_2}+\int f_1\Pi\cdot\lr{\mc{J}_1\cdot\frac{\p\log f_1}{\p\bol{z}_1}-\mc{J}_2\cdot\frac{\p\log f_2}{\p\bol{z}_2}}d\bol{z}_1\right]\right\}.\label{f2t}
\end{align}
\end{subequations}
Observe that if collisions occur among particles of the same ensemble, the evolution equation for the distribution function $f\lr{\bol{z},t}=f_1\lr{\bol{z},t}$ becomes
\begin{equation}
\frac{\p f}{\p t}=\frac{\p}{\p\bol{z}}\cdot\left\{f\mc{J}\cdot\left[-\frac{\p\lr{H+\Phi}}{\p\bol{z}}+\int f'\Pi\cdot\lr{\mc{J}'\cdot\frac{\p\log f'}{\p\bol{z}'}-\mc{J}\cdot\frac{\p\log f}{\p\bol{z}}}d\bol{z}'\right]\right\},\label{ft}
\end{equation}
where $f'=f\lr{\bol{z}',t}$, $\mc{J}'=\mc{J}\lr{\bol{z}'}$, $\Phi=\int f'V\lr{\bol{z},\bol{z}'}\,d\bol{z}'$, and $\Pi=\Pi\lr{\bol{z},\bol{z}'}$, and now we have finally derived the collision operator of equation \eqref{eq2}.

\section{Conservation laws, entropy production, and equilibria}
The aim of this section if to show that the collision operator derived in the previous section is consistent with conservation of total particle number and energy, as well as entropy growth. 
In addition, we will discuss the allowed equilibrium configurations
and their consequences for certain  systems of physical interest. 


\subsection{Conservation of particle number, energy, and interior   Casimir invariants}
Conservation of total particle numbers (total probabilities)
\begin{equation}
N_1=\int_{\Omega}f_1\,d\bol{z}_1=1,~~~~N_2=\int_{\Omega}f_2\,d\bol{z}_2=1,
\end{equation}
is a consequence of the fact that the right-hand side of equations \eqref{f1t} and \eqref{f2t} is in divergence form. Hence, assuming the phase space domain $\Omega$ to be a smooth bounded domain with boundary $\p\Omega$, the rates of change of $N_1$ and $N_2$ can be written as boundary integrals. For example,
\begin{equation}
\frac{dN_1}{dt}=\int_{\p\Omega}f_1\mc{J}_1\cdot\left[
-\frac{\p\lr{H_1+\Phi_1}}{\p \bol{z}_1}+\int_{\Omega} f_2\Pi\cdot\lr{\mc{J}_2\cdot\frac{\p\log f_2}{\p\bol{z}_2}-\mc{J}_1\cdot\frac{\p\log f_1}{\p\bol{z}_1}}\,d\bol{z}_2
\right]\cdot\bol{n}_1\,dS_1,\label{dN1dt}
\end{equation}
where $\bol{n}_1$ denotes the unit outward normal to $\p\Omega$ and $dS_1$ the corresponding surface element (if $\bol{z}_{1lab}$ are Cartesian coordinates in $\Omega$ such that $d\bol{z}_1=J_1d\bol{z}_{1lab}$ with $J_1=J\lr{\bol{z}_1}$ the Jacobian determinant at $\bol{z}_1$,  $dS_1=J_1dS_{1lab}$ where $dS_{1lab}$ is the
surface element associated with the Euclidean metric).  
The surface integral \eqref{dN1dt} vanishes under appropriate boundary conditions. For example, setting $f_1=f_2=0$ on $\p\Omega$ ensures that both $N_1$ and $N_2$ are constants (provided that $\p\log f_1/\p\bol{z}^1$ and $\p\log f_2/\p\bol{z}^2$ are well behaved on the boundary). More generally, one could demand the probability flux to be tangent to the bounding surface to ensure the thermodynamic closure of the system, i.e.
\begin{equation}
f_1\mc{J}_1\cdot\left[
-\frac{\p\lr{H_1+\Phi_1}}{\p \bol{z}_1}+\int f_2\Pi\cdot\lr{\mc{J}_2\cdot\frac{\p\log f_2}{\p\bol{z}_2}-\mc{J}_1\cdot\frac{\p\log f_1}{\p\bol{z}_1}}\,d\bol{z}_2
\right]\cdot\bol{n}_1=0~~~~{\rm on}~~\p\Omega,\label{BC}
\end{equation}
as well as a symmetric condition arising from $dN_2/dt$. 

Next, consider the total energy of the system,
\begin{equation}
\mf{H}_{12}=\int f_1f_2\mc{H}_{12}d\bol{z}_1 d\bol{z}_2=\int f_1H_1d\bol{z}_1+\int f_1f_2V\,d\bol{z}_1d\bol{z}_2+\int f_2H_2d\bol{z}_2. 
\end{equation}
Using equations \eqref{f1t} and \eqref{f2t}, we have
\begin{equation}
\begin{split}
\frac{d\mf{H}_{12}}{dt}=&\int_{\Omega}\frac{\p f_1}{\p t}\lr{H_1+\Phi_1}\,d\bol{z}_1+
\int_{\Omega}\frac{\p f_2}{\p t}\lr{H_2+\Phi_2}\,d\bol{z}_2\\
=&\int_{\p\Omega}f_{1}\lr{H_1+\Phi_1}\mc{J}_1\cdot\left[-\frac{\p\lr{H_1+\Phi_1}}{\p \bol{z}_1}+\int_{\Omega} f_2\Pi\cdot\lr{\mc{J}_2\cdot\frac{\p\log f_2}{\p\bol{z}_2}-\mc{J}_1\cdot\frac{\p\log f_1}{\p\bol{z}_1}}d\bol{z}_2\right]\cdot\bol{n}_1dS_1\\
&+\int_{\p\Omega}f_{2}\lr{H_2+\Phi_2}\mc{J}_2\cdot\left[-\frac{\p\lr{H_2+\Phi_2}}{\p \bol{z}_2}+\int_{\Omega} f_1\Pi\cdot\lr{\mc{J}_1\cdot\frac{\p\log f_1}{\p\bol{z}_1}-\mc{J}_2\cdot\frac{\p\log f_2}{\p\bol{z}_2}}d\bol{z}_1\right]\cdot\bol{n}_2dS_2\\
&+\int_{\Omega}f_1\lrs{\mc{J}_1\cdot\frac{\p\lr{H_1+\Phi_1}}{\p\bol{z}_1}}\cdot\left[\int_{\Omega}f_2\Pi\cdot\lr{\mc{J}_2\cdot\frac{\p\log f_2}{\p\bol{z}_2}-\mc{J}_1\cdot\frac{\p\log f_1}{\p\bol{z}_1}}d\bol{z}_2\right]\,d\bol{z}_1\\
&+\int_{\Omega}f_2\lrs{
\mc{J}_2\cdot\frac{\p\lr{H_2+\Phi_2}}{\p\bol{z}_2}}\cdot\left[\int_{\Omega}f_1\Pi\cdot\lr{\mc{J}_1\cdot\frac{\p\log f_1}{\p\bol{z}_1}-\mc{J}_2\cdot\frac{\p\log f_2}{\p\bol{z}_2}}d\bol{z}_1\right]
\,d\bol{z}_2.
\end{split}
\end{equation}
Boundary terms vanish under suitable boundary conditions as above. 
Then, 
\begin{equation}
\begin{split}
\frac{d\mf{H}_{12}}{dt}=&\int f_1f_2\lrs{\mc{J}_1\cdot\frac{\p\lr{H_1+\Phi_1}}{\p\bol{z}_1}}\cdot\Pi\cdot\lr{\mc{J}_2\cdot\frac{\p\log f_2}{\p\bol{z}_2}-\mc{J}_1\cdot\frac{\p\log f_1}{\p\bol{z}_1}}\,d\bol{z}_1d\bol{z}_2\\
&\int f_1f_2\lrs{\mc{J}_2\cdot\frac{\p\lr{H_2+\Phi_2}}{\p\bol{z}_2}}\cdot\Pi\cdot\lr{\mc{J}_1\cdot\frac{\p\log f_1}{\p\bol{z}_1}-\mc{J}_2\cdot\frac{\p\log f_2}{\p\bol{z}_2}}\,d\bol{z}_1d\bol{z}_2
\\
=&\frac{1}{2}\tau_c^2
\int f_1f_2\Gamma\lrs{\mc{J}_1\cdot\frac{\p\lr{H_1+\Phi_1}}{\p\bol{z}_1}}\cdot\frac{\p V}{\p\bol{z}_1}\frac{\p V}{\p\bol{z}_1}\cdot\lr{\mc{J}_2\cdot\frac{\p\log f_2}{\p\bol{z}_2}-\mc{J}_1\cdot\frac{\p\log f_1}{\p\bol{z}_1}}d\bol{z}_1d\bol{z}_2\\
&+\frac{1}{2}\tau_c^2\int f_1f_2\Gamma\lrs{\mc{J}_2\cdot\frac{\p\lr{H_2+\Phi_2}}{\p\bol{z}_2}}\cdot\frac{\p V}{\p\bol{z}_1}\frac{\p V}{\p\bol{z}_1}\cdot \lr{\mc{J}_1\cdot\frac{\p\log f_1}{\p\bol{z}_1}-\mc{J}_2\cdot\frac{\p\log f_2}{\p\bol{z}_2}}d\bol{z}_1d\bol{z}_2\\
=&\frac{1}{2}\tau_c^2\int f_1f_2\Gamma\left[\mc{J}_1\cdot\frac{\p\lr{H_1+\Phi_1}}{\p\bol{z}_1}-\mc{J}_2\cdot\frac{\p\lr{H_2+\Phi_2}}{\p\bol{z}_2}\right]\cdot\frac{\p V}{\p\bol{z}_1}\frac{\p V}{\p\bol{z}_1}\cdot\lr{\mc{J}_2\cdot\frac{\p\log f_2}{\p\bol{z}_2}-\mc{J}_1\cdot\frac{\p\log f_1}{\p\bol{z}_1}}d\bol{z}_1d\bol{z}_2\\=&0,
\end{split}
\end{equation}
where in the last passage we used the fact that energy is conserved during a collision event according to equation \eqref{dEdt3}. 
Hence, the energy $\mf{H}_{12}$ is constant. 
It is worth observing that when collisions occur between particles of the same ensemble, the functional $\mf{H}_{12}$ becomes
\begin{equation}
\mf{H}_{12}=2\int fHd\bol{z}+\int f\Phi\,d\bol{z}.
\end{equation}
Hence, the conserved energy is
\begin{equation}
\mf{H}=\int f\lr{H+\frac{1}{2}\Phi}d\bol{z}.
\end{equation}

In a similar fashion, one can verify that the interior Casimir invariant 
\begin{equation}
\mf{C}_{12}=\int f_1C_1\,d\bol{z}_1+\int f_2C_2\,d\bol{z}_2,
\end{equation}
where $C_1=C\lr{\bol{z}_1}$ and $C_2=C\lr{\bol{z}_2}$ are Casimir invariants such that $\mc{J}_1\p_{\bol{z}_1}C_1=\bol{0}$ and $\mc{J}_2\p_{\bol{z}_2}C_2=\bol{0}$, is a constant of motion. 
Indeed, assuming again boundary integrals to vanish, we have
\begin{equation}
\begin{split}
\frac{d\mf{C}_{12}}{dt}=&\int f_1\lr{\mc{J}_1\cdot\frac{\p C_1}{\p\bol{z}_1}}\cdot\lrs{-\frac{\p\lr{H_1+\Phi_1}}{\p\bol{z}_1}+\int f_2\Pi\cdot\lr{\mc{J}_2\cdot\frac{\p\log f_2}{\p\bol{z}_2}-\mc{J}_1\cdot\frac{\p\log f_1}{\p\bol{z}_1}}d\bol{z}_2}d\bol{z}_1\\ 
&+
\int f_2\lr{\mc{J}_2\cdot\frac{\p C_2}{\p\bol{z}_2}}\cdot \lrs{-\frac{\p\lr{H_2+\Phi_2}}{\p\bol{z}_2}+\int f_1\Pi\cdot\lr{\mc{J}_1\cdot\frac{\p\log f_1}{\p\bol{z}_1}-\mc{J}_2\cdot\frac{\p\log f_2}{\p\bol{z}_2}}d\bol{z}_1}d\bol{z}_2=0.
\end{split}
\end{equation}
We stress again that the interior Casimir invariant  $\mf{C}_{12}$ is induced on the field theory by the Casimir invariants $C_1$ and $C_2$ of the microscopic Poisson tensors $\mc{J}_1$ and $\mc{J}_2$. 



\subsection{Entropy production and equilibria}
Consider the following entropy measure
\begin{equation}
S_{12}=-\int f_1\log f_1d\bol{z}_1-\int f_2\log f_2d\bol{z}_2.\label{S12}
\end{equation}
The rate of change in \eqref{S12} can be evaluated with the aid of \eqref{f1t} and \eqref{f2t}. Assuming 
boundary integrals to vanish, we have
\begin{equation}
\begin{split}
\frac{dS_{12}}{dt}=&\int 
f_1\lrc{\mc{J}_1\cdot\left[-\frac{\p\lr{H_1+\Phi_1}}{\p \bol{z}_1}+\int f_2\Pi\cdot\lr{\mc{J}_2\cdot\frac{\p\log f_2}{\p\bol{z}_2}-\mc{J}_1\cdot\frac{\p\log f_1}{\p\bol{z}_1}}d\bol{z}_2\right]}\cdot\frac{\p\log f_1}{\p\bol{z}_1}\,d\bol{z}_1
\\&+f_2\lrc{\mc{J}_2\cdot\left[-\frac{\p\lr{H_2+\Phi_2}}{\p \bol{z}_2}+\int f_1\Pi\cdot\lr{\mc{J}_1\cdot\frac{\p\log f_1}{\p\bol{z}_1}-\mc{J}_2\cdot\frac{\p\log f_2}{\p\bol{z}_2}}d\bol{z}_1\right]}\cdot\frac{\p\log f_2}{\p\bol{z}_2}\,d\bol{z}_2\\=&
\int f_1f_2\lrs{\Pi\cdot\lr{\mc{J}_2\cdot\frac{\p\log f_2}{\p\bol{z}_2}-\mc{J}_1\cdot\frac{\p\log f_1}{\p\bol{z}_1}}}\cdot\lr{\mc{J}_2\cdot\frac{\p\log f_2}{\p\bol{z}_2}-\mc{J}_1\cdot\frac{\p\log f_1}{\p\bol{z}_1}}d\bol{z}_1d\bol{z}_2\\=&\frac{1}{2}\tau_c^2\int 
f_1f_2\Gamma \lrs{\frac{\p V}{\p\bol{z}_1}\cdot\lr{{\mc{J}_2\frac{\p\log f_2}{\p\bol{z}_2}-\mc{J}_1\frac{\p\log f_1}{\p\bol{z}_1}}}}^2d\bol{z}_1d\bol{z}_2
\geq 0.\label{dSdt}
\end{split}
\end{equation}
where in the last passage we used the measure preserving property \eqref{IM2} of the Poisson tensor $\mc{J}$ to eliminate terms involving $H_1+\Phi_1$ and $H_2+\Phi_2$ (which can be written as surface integrals), the expression \eqref{Pi} of the symmetric tensor $\Pi$, and the working hypothesis that $f_1$ and $f_2$ remain non-negative functions at all times so that $f_1f_2\geq 0$. 
Equation \eqref{dSdt} implies that the functional \eqref{S12} 
is a non-decreasing function of time, thus ensuring entropy growth. 
When collisions occur between particles of the same ensemble, one can verify 
in a similar manner that the relevant entropy measure is given by
\begin{equation}
S=-\int f\log f\,d\bol{z}.
\end{equation}

Equation \eqref{dSdt} has also consequences for the allowed maximum entropy configurations (thermodynamic equilibria). 
Indeed, we must have
\begin{equation}
\lim_{t\rightarrow+\infty}\frac{dS}{dt}=0, 
\end{equation}
provided that such limit exists. 
Since the integrand in equation \eqref{dSdt} is non-negative, it follows that 
a maximum entropy configuration must satisfy
\begin{equation}
\mc{J}_2\cdot\frac{\p\log f_2}{\p\bol{z}_2}-\mc{J}_1\cdot\frac{\p\log f_1}{\p\bol{z}_1}\in {\rm ker}\lr{\Pi},\label{MaxEnt1}
\end{equation}
where ${\rm ker}\lr{\Pi}$ denotes the kernel of the covariant tensor $\Pi$.
Recalling the expression of $\Pi$, equation \eqref{Pi}, we may equivalently write
\begin{equation}
\lr{\mc{J}_2\cdot\frac{\p\log f_2}{\p\bol{z}_2}-\mc{J}_1\cdot\frac{\p\log f_1}{\p\bol{z}_1}}\cdot\frac{\p V}{\p\bol{z}_1}=0.\label{MaxEnt2}
\end{equation}
Using the elastic scattering property \eqref{dEdt3}, it readily follows that 
\begin{equation}
\log f_1=-\beta\lr{H_1+\Phi_1}+g_1\lr{\bol{C}_1},~~~~\log f_2=-\beta\lr{H_2+\Phi_2}+g_2\lr{\bol{C}_2},\label{MaxEnt3}
\end{equation}
is a maximum entropy state satisfying \eqref{MaxEnt2}. 
Here, $\beta\in\mathbb{R}$, $g_1\lr{\bol{C}_1}$ and $g_2\lr{\bol{C}_2}$ are functions of the Casimir invariants $\bol{C}=\lr{C_1,...,C_k}$  spanning the kernel of the Poisson tensor $\mc{J}$ (see equation \eqref{Cas}), and the notation $\bol{C}_1=\bol{C}\lr{\bol{z}_1}$, 
$\bol{C}_2=\bol{C}\lr{\bol{z}_2}$ has been used. 
The precise form of the functions $g_1$ and $g_2$ depends on  initial conditions (the initial values of the Casimir invariants). 
It can be verified that \eqref{MaxEnt3} is the most general solution of \eqref{MaxEnt2} for arbitrary $V$ by noting that $f_1$ and $f_2$ are independent of $\bol{z}_2$ and $\bol{z}_1$ respectively, a fact that forces $\beta$ to be a spatial constant. Here, we are assuming that energy $H+\Phi$ is the only nontrivial invariant during a collision event, but the case in which other conserved quantities exist during collisions will be discussed later. In particular, we will see that the presence of other scattering invariants results in additional contributions on the right-hand side of \eqref{MaxEnt3}. 
As in the usual Boltzmann distribution, the constant $\beta$ can be interpreted as the temperature of the system. 
The positive sign of $\beta$ can be deduced from the fact that $H$ includes the kinetic energy of the colliding particles, and therefore a negative $\beta$ would result in divergence of the distribution functions at large kinetic energies. 
Now suppose that the phase space coordinates $\bol{z}_1$ and $\bol{z}_2$ bearing the invariant measure of the system 
are connected to the laboratory frame coordinates (the coordinates used to observe the system) $\bol{z}_{1lab}$ and $\bol{z}_{2lab}$ through the Jacobian determinant $J$ according to $d\bol{z}_1=J_1d\bol{z}_{1lab}$ 
and $d\bol{z}_2=J_2d\bol{z}_{2lab}$ with $J_1=J\lr{\bol{z}_1}$ and $J_2=J\lr{\bol{z}_2}$. Then, the equilibrium distribution functions 
observed in the laboratory frame are
\begin{equation}
f_{1{ lab}}=J_1\exp\left\{-\beta\lr{H_1+\Phi_1}+g_1\lr{\bol{C}_1}\right\},~~~~f_{2{ lab}}=J_2\exp\left\{-\beta\lr{H_2+\Phi_2}+g_2\lr{\bol{C}_2}\right\}.\label{MaxEnt4}
\end{equation}
We therefore see that significant departure from standard Maxwell-Boltzmann statistics may occur for collision processes in noncanonical phase spaces through the Jacobian determinant $J$ and the functions $g_1$ and $g_2$, which arise as a consequence of the geometric properties of the Poisson tensor $\mc{J}$. 

A final remark concerns the existence of other quantities that are preserved during a collision event in addition to the total energy of the colliding particles. For example, suppose that collisions do not alter the total vertical angular momentum $\ell_{z1}+\ell_{z2}$ of the interacting particles. 
This means that a condition analogous to the elastic scattering condition \eqref{dEdt3} holds, namely
\begin{equation}
\lr{\mc{J}_1\p_{\bol{z}_1}\ell_{z1}-\mc{J}_2\p_{\bol{z}_2}{\ell_{z2}}}\cdot\p_{\bol{z}_1}V=0.\label{lz12}
\end{equation}
If the vertical angular momentum $\ell_z$ is also a constant for the  motion of each particle, i.e. $\p_{\bol{z}}\ell_z\cdot\mc{J}\p_{\bol{z}}\lr{H+\Phi}=0$ (note that here $\ell_z$ does not need to be a Casimir invariant of the Poisson tensor), then one can show that the total angular momentum $L_{z12}=\int f_1\ell_{z1}\,d\bol{z}_1+\int f_2\ell_{z2}\,d\bol{z}_2$ is a constant of motion and the equilibrium \eqref{MaxEnt4} is further generalized to
\begin{equation}
f_{1lab}=J_1\exp\left\{-\beta\lr{H_1+\Phi_1}-\gamma\ell_{z1}+g_1\lr{\bol{C}_1}\right\},~~~~f_{2lab}=J_2\exp\left\{-\beta\lr{H_2+\Phi_2}-\gamma\ell_{z2}+g_2\lr{\bol{C}_2}\right\},\label{MaxEnt5}
\end{equation}
where $\gamma$ is a real constant. 
It is worth observing that here $\log f_1$ and $\log f_2$ can only be linear functions of $\ell_{z1}$ and $\ell_{z2}$: quantities that are preserved by binary interactions  always appear linearly in the logarithm of the equilibrium distribution function. This is in contrast with Casimir invariants, which affect the distribution function according to initial conditions because their value remains constant for each particle regardless of collisions.    
 A practical example of the scenario above is given by binary gravitational collisions of massive particles whose orbits are  restricted to the plane $z=0$ of a Cartesian coordinate system and whose initial total angular momentum is $\bol{L}_{12}= L_{z12}\nabla z\neq\bol{0}$ (this system could represent stars rotating within the galactic plane). Indeed, denoting with $\mc{H}_{12}=\frac{1}{2m}\bol{p}_1^2+\frac{1}{2m}\bol{p}_2^2+V\lr{\abs{\bol{q}_1-\bol{q}_2}}$ the Hamiltonian of two interacting particles, we have
 \begin{equation}
 \begin{split}
\frac{d}{dt}\lr{\ell_{z1}+\ell_{z2}}=&\lr{\dot{\bol{q}}_1\times\bol{p}_1-\bol{q}_1\times\frac{\p \mc{H}_{12}}{\p\bol{q}_1}+\dot{\bol{q}}_2\times\bol{p}_2-\bol{q}_2\times\frac{\p \mc{H}_{12}}{\p\bol{q}_2}}\cdot\nabla z\\=&\frac{\p V}{\p\bol{q}_1}\times\lr{\bol{q}_1-\bol{q}_2}\cdot\nabla z=0,
 \end{split}
 \end{equation}
 which corresponds to the condition \eqref{lz12}.

\section{Examples}
As outlined in the introduction, the main purpose of the present theory is to capture non-trivial self-organized structures that arise when a physical system is allowed to relax over spacetime scales that retain the noncanonicality of the Poisson tensor $\mc{J}$. 
Hence, our main focus is not to generalize the type of interactions $V$ driving the thermalization of the system, but rather to understand the effect of noncanonical phase spaces on collision processes driven by standard interactions, such as the Coulomb or gravitational ones. 
In this section we therefore describe some physical systems 
that are appropriately described by the evolution equations \eqref{f1t} and \eqref{f2t} derived above, and explain how the present theory relates to the phenomena of self-organization and collisionless relaxation.

\subsection{The waterbag model for the 1D Vlasov-Poisson Equation}
Consider the one dimensional Vlasov-Poisson equation in a periodic region $\Omega=[0,2\pi]$  
for the distribution function $f\lr{p,q,t}$ of an ensemble of charged particles, 
\begin{equation}
\frac{\p f}{\p t}+p\frac{\p f}{\p q}-\frac{\p\Phi}{\p q}\frac{\p f}{\p p}=0,~~~~\frac{\p^2\Phi}{\p q^2}=-\int_{\mathbb{R}}fdp.\label{VP}
\end{equation}
In this notation, $p\in\mathbb{R}$ and $q\in\Omega$ denote the momentum and position of a charged particle with energy $E\lr{p,q,t}={p^2}/{2}+\Phi$, physical constants have been set to unity, and $\Phi\lr{q,t}$ represents the electric potential energy.  
The waterbag model \cite{PJMHopfI} arises when the distribution function $f$ is piecewise constant between pair of curves $p^{\alpha}\lr{q,t}$ and $p^{\alpha+1}\lr{q,t}$ with $\alpha\in\mathbb{Z}$, i.e.  
\begin{equation}
f\lr{p,q,t}=f_\alpha\in\mathbb{R}_{\geq 0}~~~~{\rm if}~~p^{\alpha}\leq p< p^{\alpha+1}.\label{WBf}
\end{equation}
Then, noting that $\dot{q}=\p E/\p p=p$, the dynamics of the curves $p^{\alpha}$ is described by the equations
\begin{equation}
\frac{\p p^{\alpha}}{\p t}+p^{\alpha}\frac{\p p^{\alpha}}{\p q}=-\frac{\p\Phi}{\p q},~~~~\frac{\p^2\Phi}{\p q^2}=-\sum_{\alpha}f_{\alpha}\lr{p^{\alpha+1}-p^{\alpha}}.\label{WBp}
\end{equation}
System \eqref{WBp} has a noncanonical Hamiltonian structure in terms of the Hamiltonian
\begin{equation}
\mf{H}=\sum_{\alpha}\frac{\Delta f_{\alpha}}{6}\int_{\Omega}\lr{p^{\alpha}}^3dq+\frac{1}{2}\int_{\Omega}\lr{\frac{\p\Phi}{\p q}}^2dq,\label{HWB}
\end{equation}
with $\Delta f_{\alpha}=f_{\alpha-1}-f_{\alpha}$, 
and the Poisson bracket
\begin{equation}
\lrc{F,G}_\ast=-\sum_{\alpha}\frac{1}{\Delta f_{\alpha}}\int_{\Omega}\frac{\delta F}{\delta p^{\alpha}}\frac{\p}{\p q}\frac{\delta G}{\delta p^{\alpha}}\,dq,\label{PBWB}
\end{equation}
acting on functionals $F,G$ of the functions $p^{\alpha}$. In particular, 
observing that under periodic boundary conditions $\delta\int_{\Omega}\lr{\Phi_q}^2=-2\int_{\Omega}\Phi\delta\Phi_{qq}dq$ and $\delta\Phi_{qq}=-\sum_{\alpha}\Delta f_{\alpha}\delta p^{\alpha}$, 
equation \eqref{WBp} takes the form
\begin{equation}
\frac{\p p^{\alpha}}{\p t}=\lrc{p^{\alpha},\mf{H}}_{\ast}.
\end{equation}
We refer the reader to \cite{PJMHopfI} for the derivation of equations \eqref{HWB} and \eqref{PBWB},  
and to section 7 below for the definition of the Poisson bracket $\lrc{\cdot,\cdot}_{\ast}$.  
The Poisson bracket \eqref{PBWB} admits the Casimir invariants
\begin{equation}
\mf{C}^{\alpha}=\int_{\Omega}{p^{\alpha}}\,dq,\label{CasWB}
\end{equation}
which measure the areas subtended by the curves $p^{\alpha}$. Indeed,
assuming periodic boundary conditions,  
\begin{equation}
\frac{d\mf{C}^{\alpha}}{dt}=\lrc{\mf{C}^{\alpha},G}=0~~~~\forall G.
\end{equation}
Our task is now to obtain a discrete Hamiltonian system for the Fourier coefficients $c^{\alpha}_k\lr{t}$ of the expansion 
\begin{equation}
p^{\alpha}=\sum_{k=-\infty}^{\infty}c^{\alpha}_k e^{{\rm i}kq},\label{FWBp}
\end{equation}
under the hypothesis that the curves $p^{\alpha}$ are periodic in $\Omega$. 
Since the curves $p^{\alpha}$ are real valued functions, we have $c_{k}^{\alpha}=c_{-k}^{\alpha\ast}$, where $\ast$ denotes the complex conjugate. 
Substituting the Fourier representation \eqref{FWBp} into equation \eqref{WBp}, one obtains the following system of equations
for the Fourier coefficients,
\begin{subequations}
\begin{align}
&\frac{d c_k^{\alpha}}{d t}+{\rm i}\sum_m\lr{k-m}c_{m}^{\alpha}c_{k-m}^{\alpha}=-{\rm i}\sum_{\beta}\frac{f_{\beta}}{k}\lr{c_{k}^{\beta+1}-c_{k}^{\beta}}~~~~~{\rm if}~~k\neq 0,\\
&\frac{d c_0^{\alpha}}{d t}-{\rm i}\sum_mmc_{m}^{\alpha}c_{-m}^{\alpha}=0.
\end{align}\label{WBck}
\end{subequations}
Here, we enforced the boundary condition $\Phi_q\lr{0}=0$ and 
observed that periodicity implies that 
$\Phi_q=\sum_{\alpha,k\neq 0}{\rm i}k^{-1}\Delta f_{\alpha}c_k^{\alpha}\lr{e^{{\rm i}kq}-1}-q\sum_{\alpha}\Delta f_{\alpha}c_0^{\alpha}$ which gives $0=\Phi_q\lr{2\pi}=-2\pi\sum_{\alpha}\Delta f_{\alpha}c_0^{\alpha}$ and $\int_0^{2\pi}\Phi_q\,dq=-2\pi\sum_{\alpha,k\neq 0}{\rm i}k^{-1}\Delta f_{\alpha}{c_k^{\alpha}}$ so that $\Phi_q=\sum_{\alpha,k\neq 0}{\rm i}k^{-1}\Delta f_{\alpha}c_k^{\alpha}e^{{\rm i}kq}$.
On the other hand, 
\begin{equation}
0=\int_0^{2\pi}\frac{\p p^{\alpha}}{\p q}\lr{p^{\alpha}}^2dq=\sum_{m,n,k}{\rm i }m\int_0^{2\pi}c_{m}^{\alpha}c_{n}^{\alpha}c_{k}^{\alpha}\exp\lrc{{\rm i}\lr{m+n+k}q}\,dq=2\pi{\rm i}\sum_{m,n} mc_{m}^{\alpha}c_{n}^{\alpha}c_{-m-n}^{\alpha}.
\end{equation}
It follows that
\begin{equation}
\begin{split}
0=&\sum_{m,n}m\frac{\p}{\p c_{k}^{\alpha}}\lr{c_{m}^{\alpha}c_{n}^{\alpha}c_{-m-n}^{\alpha}}=\sum_{m,n}\lr{m\delta_{km}c_{n}^{\alpha}c_{-m-n}^{\alpha}+m\delta_{kn}c_{m}^{\alpha}c^{\alpha}_{-m-n}+m\delta_{k,-m-n}c_m^{\alpha}c_n^{\alpha}}\\=&
\sum_m\lr{kc_n^{\alpha}c_{-k-n}^{\alpha}+
mc_{m}^{\alpha}c_{-k-m}^{\alpha}+mc_{m}^{\alpha}c_{-k-m}^{\alpha}}=\sum_m\lr{2m+k}c_{m}^{\alpha}c_{-k-m}^{\alpha}.
\end{split}
\end{equation}
From this equation, one finds that
\begin{equation}
\sum_mkc_m^{\alpha}c_{k-m}^{\alpha}=\sum_m2mc_m^{\alpha}c_{k-m}^{\alpha},~~~~0=\sum_mmc_m^{\alpha}c_{-m}^{\alpha}.
\end{equation}
The system of equations \eqref{WBck} can therefore be expressed as
\begin{subequations}
\begin{align}
\frac{d c_{k}^{\alpha}}{d t}=&-\frac{1}{2}{\rm i}\sum_mkc_m^{\alpha}c_{k-m}^{\alpha}-{\rm i}\sum_{\beta}\frac{\Delta f_{\beta}}{k}c_k^{\beta}
~~~~{\rm if}~~k\neq0\\
\frac{d c_0^{\alpha}}{d t}=&0.
\end{align}\label{WBck2}
\end{subequations}
On the other hand, observe that the Hamiltonian \eqref{HWB} becomes
\begin{equation}
\begin{split}
\mf{H}=&\sum_{\alpha,m,n}\frac{\Delta f_{\alpha}}{3}\pi c^{\alpha}_mc_{n}^{\alpha}c_{-m-n}^{\alpha}+\frac{1}{2}\int_\Omega \abs{\sum_{\alpha,m\neq 0}\frac{\Delta f_{\alpha}}{m}c_m^{\alpha}e^{{\rm i}mq}}^2dq\\=&
\sum_{\alpha,m,n}\frac{\Delta f_{\alpha}}{3}\pi c_m^{\alpha}c_n^{\alpha}c_{-m-n}^{\alpha}+\pi\sum_{\alpha,\beta,m\neq 0}\frac{\Delta f_{\alpha}\Delta f_{\beta}}{m^2}c_m^{\alpha}c_{-m}^{\beta}.
\end{split}\label{WBH}
\end{equation}
For $k\neq 0$ we may therefore evaluate
\begin{equation}
\begin{split}
\frac{\p\mf{H}}{\p c^{\gamma}_{k}}=&\sum_{m,n}\frac{\Delta f_{\gamma}}{3}\pi\lr{2\delta_{mk}c_n^{\gamma}c_{-m-n}^{\gamma}
+\delta_{-m-n,k}c_m^{\gamma}c_n^{\gamma}}+2\pi\sum_{\beta} \frac{\Delta f_{\gamma}\Delta f_{\beta}}{k^2}c_{-k}^{\beta}\\=&
\pi\Delta f_{\gamma}\sum_{m}c_m^{\gamma}c_{-k-m}^{\gamma}+2\pi\Delta f_{\gamma} \sum_{\beta}\frac{\Delta f_{\beta}}{k^2}c_{-k}^{\beta}
.
\end{split}
\end{equation}
Similarly, 
\begin{equation}
\frac{\p\mf{H}}{\p c^{\gamma}_0}=\pi\Delta f_{\gamma}\sum_mc_m^{\gamma}c_{-m}^{\gamma}.
\end{equation}
Introducing the Poisson bracket,
\begin{equation}
\lrc{F,G}'=-\sum_{\alpha,k}
\frac{{\rm i}k}{2\pi\Delta f_{\alpha}}\frac{\p F}{\p c_k^{\alpha}}\frac{\p G}{\p c_{-k}^{\alpha}}=-\sum_{\alpha}\sum_{k=0}^{\infty}\frac{{\rm i}k}{2\pi\Delta f_{\alpha}}\lr{\frac{\p F}{\p c_{k}^{\alpha}}\frac{\p G}{\p c_{-k}^{\alpha}}-\frac{\p F}{\p c_{-k}^{\alpha}}\frac{\p G}{\p c_{k}^{\alpha}}},
\end{equation}
where $F$ and $G$ are functions of the Fourier coefficients, 
for all $k$ the equations of motion \eqref{WBck2} can be written as
\begin{equation}
\frac{d c_k^{\alpha}}{d  t}=\lrc{c_k^{\alpha},\mf{H}}'=-\frac{{\rm i}k}{2\pi\Delta f_{\alpha}}\frac{\p\mf{H}}{\p c_{-k}^{\alpha}}.\label{WBckIII}
\end{equation}
Notice that the Casimir invariants \eqref{CasWB}, which evaluate to $\mf{C}^{\alpha}=2\pi c^{\alpha}_0$, satisfy
\begin{equation}
\frac{d\mf{C}^{\alpha}}{dt}=2\pi\lrc{c_0^{\alpha},\mf{H}}'=0,
\end{equation}
for any $\mf{H}$.

Now consider the dynamical variables $\bol{c}^{\alpha}=\lr{...,c_{-2}^{\alpha},c_{-1}^{\alpha},c_0^{\alpha},c_1^{\alpha},c_2^{\alpha},...}$
and define the contravariant 2-tensor $\mc{J}^{ij}_{\alpha}$ whose non-zero components are given by
\begin{equation}
\mc{J}^{k-k}_{\alpha}=-\mc{J}^{-kk}_{\alpha}=-\frac{{\rm i}k}{2\pi\Delta f_{\alpha}}.
\end{equation}
The equations of motion \eqref{WBckIII} take the form
\begin{equation}
\frac{d c_i^{\alpha}}{d t}=\mc{J}^{ij}_{\alpha}\frac{\p\mf{H}^{\alpha}}{\p c^{\alpha}_j},
\end{equation}
where $\mf{H}^{\alpha}$ is the energy of the $\alpha$th waterbag with expression
\begin{equation}
\mf{H}^{\alpha}=\sum_{m,n}\frac{\Delta f_{\alpha}}{3}\pi c_m^{\alpha}c_{n}^{\alpha}c_{-m-n}^{\alpha}+2\pi\sum_{\beta\neq\alpha,m\neq 0}\frac{\Delta f_{\alpha}\Delta f_{\beta}}{m^2}c_m^{\alpha}c_{-m}^{\beta}+\pi\sum_{m\neq 0}\frac{\Delta f_{\alpha}^2}{m^2}c_m^{\alpha}c_{-m}^{\alpha}.
\end{equation}
Furthermore, the volume elements $dc_k^{\alpha} dc_{-k}^{\alpha}$, $k\neq 0$,  
are invariant because
\begin{equation}
\frac{\p}{\p c_k^{\alpha}}\frac{d c_{k}^{\alpha}}{d t}+\frac{\p}{\p c_{-k}^{\alpha}}\frac{d c_{-k}^{\alpha}}{d t}=0.
\end{equation}
We are now ready to apply the  theory developed in sections 2, 3, and 4. 
In particular, we consider 
the following scenario: 
the waterbags numbered by $\alpha$ 
are such that $f$ defines a zebra-like pattern with empty regions ($f_{\alpha}=0$) separating 
populated regions ($f_{\alpha}>0$). 
Two non-empty regions interact, e.g. via the Coulomb force, when some portion of their boundaries becomes sufficiently close during waterbag dynamics (this condition implies that the relative velocity of the colliding waterbags is not too large). 
However, the interacting regions do not merge, resulting in area preserving scatterings between waterbags.  
Let $\mc{F}=\mc{F}\lr{\bol{c}^{{\alpha}}}$ denote the distribution function in the space of the Fourier coefficients.  
The distribution $\mc{F}$ will therefore obey the evolution equation \eqref{ft} with 
$H+\Phi=\mf{H}^{\alpha}$ 
and with $V$ modeling binary interactions between waterbags in terms of the Fourier coefficients. The resulting equilibrium predicted by \eqref{MaxEnt3} takes the form
\begin{equation}
\mc{F}_{\infty}\propto \exp\lrs{-\beta\mf{H}^{\alpha}-g\lr{c_0^{{\alpha}}}}.
\end{equation}
This equation informs us about the effect of waterbag area preservation, which is quantified by the function $g\lr{c_0^{\alpha}}$, on the final distribution of Fourier coefficients, which determines the most probable (or average) contour profile  $\bar{p}\lr{q}=\bar{c}_k e^{{\rm i}kq}$ with $\bar{c}_k=\int\mc{F}_{\infty}c^{\alpha}_k d\bol{c}^{\alpha}$ once the system has relaxed (since $\mc{F}$ is a real valued function, the value of this integral can be evaluated by noting that $dc_k^{\alpha}\w dc_{-k}^{\alpha}=2{\rm i}A_k^{\alpha}d\varphi_k^{\alpha}\w dA_k^{\alpha}$ with $c_k^{\alpha}=A_k^{\alpha}e^{{\rm i}\varphi_k^{\alpha}}$ and $A_k^{\alpha}$ and $\varphi_k^{\alpha}$ real valued functions. 


\subsection{Charged particle diffusion by $\bol{E}\cp\bol{B}$ drift in an integrable magnetic field}
The $\bol{E}\cp\bol{B}$ drift equations of motion  
\begin{equation}
\dot{\bol{x}}=\frac{\bol{E}\cp\bol{B}}{B^2}=\frac{\bol{B}}{B^2}\cp\nabla\Phi,\label{vE}
\end{equation}
where $\bol{E}=-\nabla\Phi$ is the electric field and $\bol{z}=\bol{x}=\lr{x,y,z}$, represent 
a reduced set of equations that describe plasma dynamics in a regime in which
inertial effects are negligible due to the smallness of the mass to charge ratio, 
and the temperature of charged particles is low (the plasma is cold). 
Indeed, in the absence of additional forces, particles are only allowed 
to drift across the magnetic field with velocity \eqref{vE} 
due to force balance $\dot{\bol{x}}\cp\bol{B}+\bol{E}=\bol{0}$. 

The $\bol{E}\cp\bol{B}$ drift dynamics described by equation \eqref{vE} is at the basis of  turbulence and transport processes in magnetically confined plasmas, 
such as drift wave turbulence in tokamaks \cite{Hasegawa} 
or inward diffusion in magnetospheres \cite{NSAIP}. 
Furthermore, the mathematical structure of equation \eqref{vE} 
generalizes rigid body dynamics \cite{Sato2016,SatoPRE2}. Indeed, 
by identifying $\bol{x}$ with the components of the angular momentum of a rigid body, 
choosing $\Phi=\lr{x^2I^{-1}_x+y^2 I^{-1}_y+z^2I^{-1}_z}/2$ to represent the rigid body energy, with $I_x$, $I_y$, and $I_z$ denoting the moments of inertia, and setting $\bol{B}=\bol{x}/\bol{x}^2$, one obtains the rigid body equations of motion. 

It is well known \cite{SatoPRE2,Chandre} that a necessary condition for equation \eqref{vE} to define an Hamiltonian system is that the candidate Poisson tensor $B^{-2}\bol{B}\cp$ satisfies the Jacobi identity 
\begin{equation}
\bol{B}\cdot\nabla\cp\bol{B}=0.\label{BdB}
\end{equation}
The vanishing of helicity density expressed by equation \eqref{BdB} 
is nothing but the Frobenius integrability condition for the magnetic field: 
if \eqref{BdB} holds there exist local functions $\lambda\lr{\bol{x}}$ and $C\lr{\bol{x}}$ such that 
$\bol{B}=\lambda\nabla C$. 
Furthermore, the function $C$ is a Casimir invariant because $\dot{C}=\dot{\bol{x}}\cdot\nabla C=0$ for any $\Phi$. 
Equation \eqref{BdB} is also a necessary condition for the existence of an invariant measure independent of the Hamiltonian $\Phi$. 
The corresponding preserved phase space volume is $dV_{I}=\lambda\abs{\nabla C}^2d\bol{x}$. 

In this subsection we consider an ensemble of 
charged particles obeying \eqref{vE} where $\Phi=\int Vf'd\bol{x}'$ denotes the ensemble averaged interaction energy and $f\lr{\bol{x},t}$ the particle distribution function in some smooth bounded domain $\Omega\subset\mathbb{R}^3$. These particles are also allowed to interact through binary collisions with potential energy $V$ (e.g. the Coulomb potential energy).
We restrict our attention to configurations in which the magnetic field defines the normal of a two dimensional surface, i.e. $\bol{B}=\lambda\nabla C\neq\bol{0}$ with $\lambda >0$, 
since this conditions guarantees the existence of a globally defined invariant measure $dV_I$ required for the formulation of the collision operator discussed in sections 2 and 3. Let $g\lr{\bol{x},t}$ denote the distribution function on the invariant measure $dV_I$. Since $fd\bol{x}=gdV_{I}$, it follows that $f=g\lambda\abs{\nabla C}^2$.  
The evolution equation for the distribution function $g$ can therefore be written in terms of $f$ with the aid of equation \eqref{ft},
\begin{equation}
\begin{split}
\frac{\p f}{\p t}=&-\nabla\cdot\lr{f\frac{\nabla C\cp\nabla\Phi}{\lambda\abs{\nabla C}^2}}+\\
&\nabla\cdot\lrc{f\frac{\nabla C}{\lambda\abs{\nabla C}^2}\times{\int f'\Pi\cdot\lrs{\frac{\nabla C'}{\lambda'\abs{\nabla C'}^2}\cp\nabla\log\lr{\frac{f'}{\lambda'\abs{\nabla C'}^2}}-\frac{\nabla C}{\lambda\abs{\nabla C}^2}\cp\nabla\log\lr{\frac{f}{\lambda\abs{\nabla C}^2}}}d\bol{x}'}}.
\end{split}
\end{equation}
According to equations \eqref{MaxEnt3} and \eqref{MaxEnt4} the resulting equilibrium distribution function has expression
\begin{equation}
f_{\infty}=\lambda\abs{\nabla C}^2\exp\lrc{-\beta\Phi+g\lr{C}}=\frac{\bol{B}^2}{\lambda}\exp\lrc{-\beta\Phi+g\lr{C}},\label{fExB}
\end{equation}
where $g\lr{C}$ is some function of $C$ determined by initial conditions. 
From this equation, it is clear that $f_{\infty}$ exhibits a strong departure from 
a Boltzmann distribution $f_{{\rm B}}\propto \exp\lrc{-\beta \Phi}$ due to the
distortion $\lambda\abs{\nabla C}^2=\bol{B}^2/\lambda$ of the invariant measure caused by the magnetic field inhomogeneity and the conservation of the Casimir invariant $C$. 
The distribution function \eqref{fExB} thus represents a clear-cut example of self-organization caused by the noncanonical Hamiltonian structure of the phase space.  
Special cases of interest are vacuum fields ($\lambda=1$), such as a dipole magnetic field, and straight magnetic fields ($\lambda=1$, $C=B_0z$, $B_0\in\mathbb{R}$).

\subsection{Self-organization in a collisionless magnetized plasma}
The physical parameters characterizing magnetized plasmas often 
arise in combinations such that binary Coulomb interactions between charged particles can be effectively neglected. Such plasmas are `collisionless'. Examples include both laboratory experiments in which a plasma is confined with a strong magnetic field \cite{YosPRL,Boxer,Ken}, and astrophysical systems such as accretion disks or plasmas trapped in planetary magnetospheres \cite{Sche,Kawazura20}. 
There are two common features that characterize these systems.
First, equilibrium configurations may exist. This means that they are achieved over spatial and time scales where the effect of binary Coulomb collisions is negligible. 
A fundamental theoretical issue therefore exists with respect to the mechanism enabling such thermalization in the absence of a microscopic dissipation process. 
Secondly, these equilibria exhibit self-organized structures.  
In particular, the distribution functions sensibly depart from the Maxwell-Boltzmann distribution. 
At first glance, this fact may appear to be inconsistent with the second law of thermodynamics, which requires entropy to grow in a closed thermodynamic system.
Here, we emphasize that treating such systems as open thermodynamic systems is not sufficient to account for their self-organizing behavior.
In the following, we argue that the theory developed in the present paper can explain both collisionless thermalization and self-organization in consistency with the laws of thermodynamics. To see this, we shall discuss 
the self-organization of a collisionless non-neutral plasma confined by a strong  magnetic field. 

To give the calculation a practical context we will use physical parameters that are consistent with laboratory experiments involving the confinement of charged particles in a dipole magnetic field  (see for example \cite{Helander2014,Sato2023}). 
Consider an electron plasma in a static magnetic field $\bol{B}=\bol{B}\lr{\bol{x}}$ with typical field strength $B\sim 1T$ (we could mimic a laboratory or planetary magnetosphere by taking $\bol{B}$ to be a dipole magnetic field, but the argument holds true for any magnetic field). 
Let $L\sim 1\,m$ be the system size (this could be the size of the confining device) and suppose that the typical electron kinetic energy is $\beta^{-1}_e=k_BT_e\sim 10\,eV$, with $k_B$ the Boltzmann constant.  
The electrons tend to move along the magnetic field while performing 
cyclotron gyration with gyroradius
\begin{equation}
r_g=\frac{mv_{\perp}}{eB}\sim 10^{-5}m,
\end{equation}
where $v_{\perp}$ is the electron velocity perpendicular to $\bol{B}$, $m$ the electron mass, and $e>0$ the modulus of the electron charge.
Denoting with $n\sim 10^{12}\,m^{-3}$ the typical electron density, 
the distance at which binary Coulomb collisions result in energy changes comparable to the electron kinetic energy can be estimated as
\begin{equation}
r_{c}=\frac{e^2}{4\pi\epsilon_0k_BT_e}\sim 10^{-10}\,m,
\end{equation}
while the average distance among particles is
\begin{equation}
r_d=n^{-1/3}\sim 10^{-4}\,m.
\end{equation}
Furthermore, the frequency of binary Coulomb collisions can be estimated \cite{Goldston} as
\begin{equation}
\nu_c\sim\frac{e^4\log\Lambda}{12\pi^{3/2}\epsilon_0^2 m^{1/2}}n\beta^{3/2}_e\sim 6\times10^{-2}\log\Lambda\,s^{-1},
\end{equation}
where $\log\Lambda$ is the Coulomb logarithm. 
Now observe that since by hypothesis the deflection angle is small, $\chi<<1$,  the Coulomb logarithm can be approximated as
\begin{equation}
\log\Lambda=\log\lr{\frac{\chi_{\rm max}}{\chi_{\rm min}}}\sim\log\lr{\frac{\sin\chi_{\rm max}}{\sin\chi_{\rm min}}}\sim\log\lr{\frac{b_{\rm max}}{b_{\rm min}}},
\end{equation}
where $b_{\rm max}$ and $b_{\rm min}$ are the maximum and minimum impact parameters. 
The order of $b_{\rm min}$ can be estimated in terms of the distance between the two colliding particles resulting in a Coulomb energy of the same order as the average kinetic energy. 
However, the maximum impact parameter $b_{\rm max}$ cannot be directly estimated in terms of the Debye length $\lambda_D=\sqrt{\epsilon_0/\beta_e ne^2}$ because the system is not neutral. 
Nevertheless, observing that the average distance among particles is several orders of magnitude greater than the distance required for a collision to result in a significant change in the particle energies, $r_d>>r_c$, we may take $b_{\rm max}\sim r_d$ because collisions between particles at distances larger than $r_d$ will result in extremely small deflections. It follows that $\log\Lambda\sim 14$ so that we obtain a rough estimate of the collision frequency $\nu_{c}\sim 1\, s^{-1}$, implying that the typical time interval after which a particle experiences a binary Coulomb scattering is $\tau_c\sim 1\,s$. 
On the other hand, experimental evidence \cite{YosPRL,Ken} suggests that the electron plasma achieves a self-organized equilibrium state characterized by an inhomogeneous density profile after a relaxation time $\tau_r\sim 0.1\,s<\tau_c$. Hence, the present system can be characterized as `collisionless', in the sense that binary Coulomb scatterings cannot provide the dissipative mechanism by which the electron plasma reaches thermal equilibrium. 

Let us see how the theory developed in the previous sections can be applied to understand the behavior of such electron plasma. 
First, observe that in the present setting the cyclotron frequency is
\begin{equation}
\omega_c=\frac{eB}{m}\sim 10^{11}\,Hz.
\end{equation}
We know from experiments \cite{YosPRL,Ken} that the spectrum of electromagnetic fluctuations
arising during the relaxation of the system  
is bounded by much lower frequencies. In particular, 
denoting with $\nu_f$ the frequency of electromagnetic fluctuations, one finds $\nu_f<<10^5\,H_z<<\omega_c$. Hence, we may safely assume that charged particles preserve their magnetic moment $\mu$, and charged particle motion is appropriately described by guiding center dynamics \cite{Cary}. We remark that this holds true because the system is collisionless, and fluctuations are much slower than the period of adiabatic motion \cite{LandauMec}.  
It is customary to use the set of coordinates  $\lr{\mu,v_{\parallel},x,y,z,\vartheta_c}$ to span the phase space of a charged particle, where $v_{\parallel}$ denotes the component of the particle velocity along the magnetic field, $\lr{x,y,z}$ Cartesian coordinates in $\mathbb{R}^3$ corresponding to the reference frame in the laboratory, and $\vartheta_c$ the phase of the cyclotron gyration. 
In this setting, the aforementioned conservation of the magnetic moment $\mu$ is associated with a Noether symmetry of the   guiding center Hamiltonian $H_{gc}$, which is independent of $\vartheta_c$, i.e. $\dot{\mu}\propto-\p_{\vartheta_c}H_{gc}=0$. 
The phase $\vartheta_c$ can thus be eliminated from the description of the system, resulting in a reduced $5$-dimensional phase space $\bol{z}=\lr{\mu,v_{\parallel},x,y,z}$. The corresponding Poisson tensor has the following form \cite{Cary}, 
\begin{equation}
\mc{J}_{gc}=\begin{bmatrix}
0&\bol{0}_{14}\\
\bol{0}_{41}&\mc{J}_{gc}^{\mu},\label{Jgc}
\end{bmatrix}
\end{equation}
Here, 
$\bol{0}_{i j}$ is the null matrix of dimension $i\times j$ and 
$\mc{J}_{gc}^{\mu}$ a $4$-dimensional matrix. 
Observe that $\mu$ is a Casimir invariant, i.e.
$\mc{J}_{gc}\p_{\bol{z}}\mu=\bol{0}$. 
Furthermore, the invariant measure (preserved phase space volume) is given by
\begin{equation}
d\Omega=B\,d\bol{z}=B\,d\mu dv_{\parallel} dxdydz.\label{IMgc}
\end{equation}
More precisely, when the magnetic field $\bol{B}$ has non-vanishing helicity density $\bol{B}\cdot\nabla\times\bol{B}\neq 0$, the above formula must be corrected by a factor depending on such helicity density. Nevertheless, we shall omit this correction to simplify the exposition, and observe that the same logic applies in either case. We also remark that a dipole magnetic field is a vacuum field $\nabla\times\bol{B}=\bol{0}$ outside the current loop generating it, and therefore the helicity density vanishes in the region occupied by the plasma. The formula \eqref{IMgc} is therefore exact for the case of a self-organizing plasma within a planetary magnetosphere. 

Now consider a cluster of $N$ electrons performing guiding center dynamics $\dot{\bol{z}}=\mc{J}_{gc}\p_{\bol{z}}H_{gc}$ within the magnetic field $\bol{B}$, and suppose that they share roughly the same value of magnetic moment $\mu$, parallel velocity $v_{\parallel}$, and spatial position $\lr{x,y,z}$ (we will make this statement more precise later). We may regard such cluster of particles as a mesoscopic particle with mass $Nm$, charge $-Ne$, magnetic moment $N\mu$, and energy $NH_{gc}$, located at the position of the center of mass. The idea is now to consider binary Coulomb collision of clusters of electrons. 
Indeed, if the cluster size $r_N$ is large enough, we expect the collision frequency to grow accordingly. Then,  collisions of electron clusters can provide a fast enough dissipative mechanism to thermalize the system. 
For this construction to hold, the number of particles contained within a sphere of radius $r_N$ must be large, i.e. $r_N>>r_d$. At the same time, these clusters must be smaller than the system size, $r_N<<L$. In the present setting, these conditions can be satisfied by choosing $r_N\sim 10^{-2}\,m$. Then, the number of particles contained in a sphere of radius $r_N$ is $N=4\pi r_N^3 n/3\sim 4\times10^6>>1$. This number takes into account electrons having different magnetic momenta and parallel velocities. However, we may assume that the system has been initially rearranged so that each cluster contains electrons with roughly the same  magnetic moment and parallel velocity (otherwise, one could simply consider a small fraction of $N$, such as $N'=10^{-3}N\sim 4\times10^{3}$, as representative electron number for an electron cluster).  
Furthermore, the distance between the centers of mass of two clusters of electrons  at which binary Coulomb collisions result in a significant change in energy is 
\begin{equation}
r_{Nc}=\frac{N^2e^2}{4\pi\epsilon_0Nk_BT_e}=Nr_c\sim 4\times 10^{-4}m.
\end{equation}
Note that, since $r_N>>r_{Nc}$, this implies that for such large deflections to occur the two clusters must penetrate each other. 
Next, the frequency of collisions can be estimated as in the case of electron-electron collisions, 
\begin{equation}
\nu_{Nc}\sim \frac{N^4e^4\log\Lambda_N}{12\pi^{3/2}\epsilon_0^{2}N^{1/2}m^{1/2}}\frac{n}{N}N^{-3/2}\beta^{3/2}_e=N\nu_c\sim 2\times 10^5\log\Lambda_N\,s^{-1},
\end{equation}
where the Coulomb logarithm for the process can be approximated as  $\log\Lambda_N\sim\log\lr{r_{Nd}/r_{Nc}}\sim 4$, with $r_{Nd}=\lr{n/N}^{-1/3}\sim 1.6\times 10^{-2}\,m>>r_{Nc}$ the typical cluster distance. We therefore obtain a rough estimate for the collision frequency, $\nu_{Nc}\sim  10^6\,s^{-1}$, and a collision time $\tau_{Nc}\sim 10^{-6}\, s$, which is now much shorter than the relaxation time $\tau_r\sim 0.1\,s$.    

Let $f\lr{\mu,v_{\parallel},x,y,z,t}$ denote the distribution function of electron clusters defined with respect to the invariant measure (preserved phase space volume) \eqref{IMgc}. The setting described above suggests that the appropriate evolution equation for the system is given by equation \eqref{ft} where the interaction tensor $\Pi$ is that associated with the Coulomb potential energy $V=N^2e^2/4\pi\epsilon_0\abs{\bol{q}_1-\bol{q}_2}$, with $\bol{q}=\lr{x,y,z}$, 
while the Poisson tensor is that associated with guiding center dynamics, i.e. the operator $\mc{J}_{gc}$ of equation \eqref{Jgc} expressed in a coordinate system with volume element \eqref{IMgc}. For example, one can use the set of coordinates $\bol{z}'=\lr{\mu,Bv_{\parallel},x,y,z}$, or any suitable magnetic coordinate system such as $\bol{z}'=\lr{\mu,v_{\parallel},\ell,\Psi,\theta}$ where $\ell$ is a length coordinate along the magnetic field (provided that it exists), $\Psi$ the flux function, and $\theta$ an angle variable such that $\bol{B}=\nabla\Psi\times\nabla\theta$. The expression of $\Pi$ for the Coulomb interaction will be discussed in more detail in section 6, where the relationship between the present theory and the Landau model is discussed. 
What is important here is the form of thermal equilibrium predicted by equation \eqref{MaxEnt3}. Indeed, recalling that $\mu$ is a Casimir invariant, 
the equilibrium distribution function becomes
\begin{equation}
f=\exp\lrc{-\beta NH_{gc}+g\lr{\mu}}.\label{feq1}
\end{equation}
The guiding center Hamiltonian can be written as $H_{gc}=\frac{1}{2}mv_{\parallel}^2+\mu B-e\Phi-\frac{1}{2}m\bol{v}_{\bol{E}}^2$, where $\Phi$ is the electric potential and $\bol{v}_{\bol{E}}=\bol{B}\times\nabla\Phi/B^2$ the $\bol{E}\times\bol{B}$ velocity. 
On the other hand, the form of the function $g\lr{\mu}$ will generally depend on the initial distribution of magnetic moment. Nevertheless, it is useful to examine the linear case $g=-\log Z-\gamma N\mu$, with $Z$ and $\gamma$ positive real constants.
This choice can be interpreted in two ways: small departure from Maxwell-Boltzmann statistics, where the function $g$ is expanded in powers of the dimensionless quantity $\gamma N\mu$, or a grand canonical ensemble (see e.g. \cite{Yos2014}) in which the equilibrium state is obtained by maximizing entropy $S=-\int f\log f\,d\Omega$ under the constraint that total particle number $N_{\rm tot}=\int fN\,d\Omega$, total energy $\mf{H}=\int fNH_{gc}\,d\Omega$, and total magnetic moment $M=\int fN\mu\,d\Omega$ are constant, according to the variational principle $\delta\lr{S-\alpha N_{\rm tot}-\beta \mf{H}-\gamma M}$ with Lagrange multipliers $\alpha$, $\beta$, and $\gamma$ such that $Z=e^{1+\alpha N}$. In either case, the equilibrium distribution function \eqref{feq1} becomes
\begin{equation}
f=\frac{1}{Z}\exp\lrc{-\beta N\lrs{\frac{1}{2}m\lr{v^2_{\parallel}-\bol{v}_{\bol{E}}^2}-e\Phi}-\mu N\lr{\beta B+\gamma}}.
\end{equation}
Recalling \eqref{IMgc}, the corresponding electron cluster spatial density $\rho_{\rm lab}\lr{x,y,z}$ in the laboratory frame can therefore be evaluated as
\begin{equation}
\rho_{\rm lab}=\int_{\mathbb{R}}dv_{\parallel}\int_0^{+\infty}d\mu f B=\frac{1}{Z}\sqrt{\frac{2\pi}{\beta N^3m}}\exp\lrc{\beta N\lrs{\frac{1}{2}m\bol{v}_{\bol{E}}^2+e\Phi}}\frac{B}{\beta B+\gamma}.
\end{equation}
Since the electron cluster temperature $\beta^{-1}$ is expected to be related to the electron temperature $\beta_e^{-1}$ according to $\beta_e=N\beta$, and observing that $\rho_{\rm lab}$ is normalized to unity through the constant $Z$, we arrive at the 
equilibrium electron spatial density in the Cartesian reference frame $\lr{x,y,z}$
\begin{equation}
\rho_{\rm lab}^{e}=\frac{1}{Z}\sqrt{\frac{2\pi}{\beta_e m}}\exp\lrc{\beta_e\lrs{\frac{1}{2}m\bol{v}_{\bol{E}}^2+e\Phi}}\frac{B}{\beta_e B+N\gamma}.\label{rhoe}
\end{equation}
In addition to the usual Boltzman factor, the density \eqref{rhoe} exhibits spatial inhomogeneity due to the conservation of the magnetic moment, which is expressed by the fact that $\gamma\neq 0$. In particular, $\rho_{\rm lab}$ will be higher in regions where the modulus $B$ is stronger. 
Other adiabatic invariants, which would appear as additional Casimir invariants of the Poisson tensor, would introduce additional structure to the equilibrium distribution function and the corresponding laboratory density. 
For example, it is known \cite{Yos2014,Sato2016,Sato2023} that in the context of planetary magnetospheres where $\bol{B}$ is essentially a dipole magnetic field, adding conservation of bounce action $j_{\parallel}$ to the above model results in concentration of charged particles along the equator with corresponding self-organization of radiation belt-type structures. 

We conclude this part by observing that the self-organized density profile \eqref{rhoe} cannot be obtained as a steady solution of the Vlasov model or the Landau kinetic equation for an ensemble of charged particles. 
Indeed, in both cases the distribution function is restricted to a function of the charged particle Hamiltonian, $f=f\lr{\frac{1}{2}m\bol{v}^2-e\Phi}$. 
This example therefore shows that it is crucial to embed the noncanonicality of a system into its kinetic equation in order to appropriately describe  relaxation processes occurring  over shorter time scales than those of binary collisions.

\subsection{Relaxation of a collisionless system of self-gravitating bodies}

A similar construction to the one obtained for a self-organizing collisionless magnetized plasma can be applied to a system of self-gravitating bodies that are separated by such large distances that binary graviatational interactions do not occur over a given time scale of interest. This situation is often encountered in astrophysics. 
A well known example is the formation of galactic structures, where the stellar distribution suggests the existence of an equilibrium state that cannot be however explained in terms of binary collisions \cite{Lynden}.

A possible approach to understand the creation of such equilibria is to exploit the Vlasov equation, because binary interactions are so unlikely that the correlation term $f_1f_2-f_{12}$ on the right-hand side of the first equation of the BBGKY hierarchy \eqref{BBGKY1} can be neglected. Then, the dynamics of the system is driven only by the collective gravitational potential $\Phi$ according to the Vlasov equation.  
In this context, equilibria are understood in terms of the coarse-grained distribution function, i.e. the distribution function resulting from an averaging process of the fine-grained distribution function (which obeys the gravitational Vlasov equation) reflecting the finite resolution used to resolve the phase space \cite{Lynden}.  

We suggest that an accurate kinetic description of collisionless relaxation of self-gravitating bodies can be obtained through the derived equation \eqref{ft}, where $f$ denotes the distribution of clusters of such bodies. 
To see this, consider the case of an ensemble of stars with typical mass $M_{\odot}=2\times 10^{30}\,kg$ and average velocity $v_{\odot}\sim 2\times 10^{4}\,m\,s^{-1}$ gravitating within a galaxy (see \cite{Maoz} for details on the values of physical parameters used here and in the paragraphs below). 
The distance at which binary gravitational collisions result in energy changes comparable with the kinetic energy of a star is
\begin{equation}
r_c=\frac{2GM_{\odot}}{v_{\odot}^2}\sim 7\times 10^{11}\,m.
\end{equation}
On the other hand, the typical distance among stars in a galaxy is $r_{d}\sim n_{\odot}^{-1/3}\sim 3\times 10^{16}\,m>>r_c$, where $n_{\odot}\sim 3\times 10^{-50}\,m^{-3}$ is the star density. 
Furthermore, the frequency of binary gravitational collisions can be estimated as
\begin{equation}
\nu_c= n_{\odot}\sigma v_{\odot}\sim 0.5\times 10^{-23}\,s^{-1},
\end{equation}
where $\sigma\sim 10^{22}\,m^2$ is an effective scattering cross-section. 

Next, consider a cluster of stars in a spherical region of radius $r_N\sim 10^{18}\,m$, with $r_c<<r_d<<r_N<<L$ where $L\sim 10^{21}\,m$ is the spatial scale (radial size) of the galaxy. The number of stars contained in a cluster is $N=4\pi r_N^3n_{\odot}/3\sim 1.2\times 10^5$. 
Now the distance at which the star cluster kinetic energy is significantly modified by a gravitational interaction with another cluster of stars 
is $r_{Nc}=Nr_c\sim 8\times 10^{16}\,m$. In addition, a lower bound for the frequency of gravitational collisions 
between star clusters can then be estimated by using the cluster physical section as scattering cross-section. We have
\begin{equation}
\nu_{Nc}\geq\frac{n_{\odot}}{N}\pi \lr{2r_{N}}^2v_{\odot}\sim 10^{-13}\,s^{-1},
\end{equation}
which gives a time interval between collisions $\tau_{Nc}\sim 10^{13}s$. 
This value is much smaller than the age of the universe $\tau_r\sim 0.4\times 10^{18}\,s$, which can be taken as the order of magnitude of the relaxation time for the system.  
Denoting with $f$ the distribution function of star clusters in the phase space (usually the invariant measure is conveniently expressed as  $d\bol{z}=r^2 dv_rdrd\omega d\varphi dv_z dz$, where $\lr{r,\varphi,z}$ denote cylindrical coordinates and $v_r$, $\omega$, and $v_z$ represent radial, angular, and vertical velocities), we may therefore represent the evolution of $f$ with the aid of equation \eqref{ft} where 
$\Pi$ is the interaction tensor for the gravitational force and the Poisson tensor 
$\mc{J}_{\rm star}$ is tailored for the specific dynamics under consideration. 
For example, if star dynamics is restricted to the galactic plane $z=0$, we expect the distance from the galactic plane $\abs{z}$ to be a Casimir invariant such that $\mc{J}_{\rm star}\p_{\bol{z}} \abs{z}=\bol{0}$.  
Similarly, if the total vertical angular momentum $L_z=N\int f\ell_z\,d\bol{z}$, with $\ell_z=M_{\odot}\omega r^2$, is preserved by the relaxation process (this is the case of binary gravitational encounters on the plane $z=0$), the interaction tensor associated with the right-hand side of equation \eqref{ft} possesses a corresponding kernel (recall equations \eqref{lz12} and \eqref{MaxEnt5}). These considerations therefore lead to equilibrium solutions of equation \eqref{ft} of the type
\begin{equation}
f=\frac{1}{Z}\exp\lrc{-\beta N H_{\odot}-\gamma N \abs{z}-\eta N \ell_z},
\end{equation}
with corresponding spatial star cluster density $\rho_{ lab}\lr{x,y,z}$ in the Cartesian  frame $\lr{x,y,z}$ given by
\begin{equation}
\rho_{lab}=\frac{1}{Z}\lr{\frac{2\pi}{\beta NM_{\odot}}}^{3/2} \exp\lrc{-\beta N\Phi-\gamma N\abs{z}+\frac{\eta^2 NM_{\odot}}{2\beta}r^2}
,\label{rhostar}
\end{equation}
where we used the expression for the energy of a star $H_{\rm \odot}=\frac{1}{2}M_{\odot}\bol{v}^2+\Phi$, with $\Phi$ the gravitational potential energy. 
We note that, due to the attractive nature of the gravitational force, one expects that  $\lim_{r\rightarrow +\infty}\rho_{ lab}=0$. 
This implies that $\Phi$, which is determined by the Poisson equation $\Delta\Phi=4\pi GM_{\odot}\lr{NM_{\odot}N_{\rm cl}\rho_{ lab}+s_{ext}}$, with $G$ the gravitational constant, $N_{\rm cl}$ the total number of clusters in the ensemble, and $s_{ext}$ any external source of gravity (e.g. a central black hole sustaining the Keplerian rotation of the galactic disk), must cancel the divergence of the centrifugal term proportional to $\eta^2$ in \eqref{rhostar} at large radii.  
For completeness, we also emphasize that equation \eqref{rhostar} does not take into account general relativistic effects. 
Due to the large masses and high kinetic energies involved, 
these effects are expected to sensibly modify the phase space structure (in particular, the invariant measure), and thus the observed spatial density. 


\section{Binary Coulomb collisions and the Landau limit}
The purpose of this section is to show that the collision operator \eqref{C7} 
reduces to the Landau collision operator \cite{Landau1936,Kampen} when charged particles interact via binary Coulomb collisions resulting in small deflections of the particle orbits. First, recall that in the case of binary Coulomb collisions the scattering volume density per unit time $\mc{V}$ can be written in the form of equation \eqref{svc}. Substituting this expression into \eqref{C7}, and noting that $\mc{J}=\mc{J}_s$ is the symplectic matrix in canonical phase space $\lr{\bol{p},\bol{q}}$, one thus obtains
\begin{equation}
\begin{split}
\mc{C}\lr{f_1,f_2}=&\frac{\tau_c^2}{2m^6}\frac{\p}{\p\bol{p}_1}\cdot\left[\int f_1f_2\sigma\abs{\bol{v}_1-\bol{v}_2}\dot{\bol{p}}_1\dot{\bol{p}}_1\cdot\lr{\frac{\p\log f_1}{\p\bol{p_1}}-\frac{\p\log f_2}{\p\bol{p}_2}}\delta\lr{\bol{q}_1-\bol{q}_2}
d\bol{p}_1'd\bol{p}_2'd\bol{p}_2d\bol{q}_2\right],\label{CL}
\end{split}
\end{equation}
where $\bol{v}_1=\bol{p}_1/m$ and $\bol{v}_2=\bol{p}_2/m$, with $m$  
the particle mass. A further integration with respect to $\bol{q}_2$ leaves only contributions due collisions occurring at the same spatial position $\bol{q}_1=\bol{q}_2$. Hence, from this point on $f_1=f_1\lr{\bol{p}_1,\bol{q}_1,t}$ and $f_2=f_2\lr{\bol{p}_2,\bol{q}_1,t}$. 
Furthermore, due to the short time scale $\tau_c$ of the interaction, the quantity
$\tau_c\dot{\bol{p}}_1$ is nothing but the total change of the momentum $\bol{p}_1$ due to the collision, i.e. $\tau_c\dot{\bol{p}}_1=\delta\bol{p}_1=\bol{p}_1'-\bol{p}_1=m\lr{\bol{v}_1'-\bol{v}_1}=m\delta\bol{v}_1$, where we introduced the quantity $\delta\bol{v}_1=\bol{v}_1'-\bol{v}_1$. Equation \eqref{CL} therefore becomes 
\begin{equation}
\begin{split}
\mc{C}\lr{f_1,f_2}=&\frac{1}{2}\frac{\p}{\p\bol{v}_1}\cdot\left[\int f_1f_2\sigma\abs{\bol{v}_1-\bol{v}_2}\delta\bol{v}_1\delta\bol{v}_1\cdot\lr{\frac{\p\log f_1}{\p\bol{v_1}}-\frac{\p\log f_2}{\p\bol{v}_2}}
m^3d\bol{v}_1'd\bol{v}_2'd\bol{v}_2\right],\label{CL2}
\end{split}
\end{equation}
Since $f_1$ and $f_2$ are defined with respect to the phase space volumes  $d\bol{p}_1d\bol{q}_1=m^3d\bol{v}_1d\bol{q}_1$ and $d\bol{p}_2d\bol{q}_2=m^3d\bol{v}_2d\bol{q}_2$ respectively, it is convenient 
to consider the corresponding distributions in velocity space, i.e. 
from now on $f_1=f_1\lr{\bol{v}_1,\bol{q}_1}=m^{3}f_1\lr{\bol{p}_1,\bol{q}_1}$ 
and similarly for $f_2$. It readily follows that the associated collision operator 
becomes
\begin{equation}
\begin{split}
\mc{C}\lr{f_1,f_2}=&\frac{1}{2}\frac{\p}{\p\bol{v}_1}\cdot\left[\int f_1f_2\sigma\abs{\bol{v}_1-\bol{v}_2}\delta\bol{v}_1\delta\bol{v}_1\cdot\lr{\frac{\p\log f_1}{\p\bol{v_1}}-\frac{\p\log f_2}{\p\bol{v}_2}}
d\bol{v}_1'd\bol{v}_2'd\bol{v}_2\right].\label{C3}
\end{split}
\end{equation}
For a binary Coulomb collision the scattering cross-section can be 
simplified according to
\begin{equation}
\sigma\lr{\bol{v}_1,\bol{v}_2;\bol{v}_1',\bol{v}_2'}d\bol{v}_1'd\bol{v}_2'=\frac{d\sigma\lr{u,\chi}}{d\Omega}d\Omega,~~~~\frac{d\sigma}{d\Omega}=\frac{1}{4}\lr{\frac{e^2}{2\pi\epsilon_0mu^2}}^2\frac{1}{\sin^4\lr{\chi/2}},
\end{equation}
where $d\sigma/d\Omega$ is the differential scattering cross-section, $u$ the modulus of the relative velocity $\bol{u}=\bol{v}_1-\bol{v}_2$, $d\Omega=\sin\chi d\chi d\phi$, $\chi$ the deflection angle of the relative velocity $\bol{u}$ caused by the collision, $\phi$ an angle determining the plane of the collision, $e$ the particle charge, and $\epsilon_0$ the vacuum permittivity. 
Furthermore, conservation of total momentum implies that the change in relative velocity $\bol{u}$ satisfies $\delta\bol{u}=2\delta\bol{v}_1$. 
Hence, 
\begin{equation}
\begin{split}
\mc{C}\lr{f_1,f_2}=&\frac{1}{8}\lr{\frac{e^2}{4\pi\epsilon_0m}}^2\frac{\p}{\p\bol{v}_1}\cdot\left[\int f_1f_2\frac{\delta\bol{u}\delta\bol{u}}{u^3\sin^4\lr{\chi/2}}\cdot\lr{\frac{\p\log f_1}{\p\bol{v_1}}-\frac{\p\log f_2}{\p\bol{v}_2}}
d\bol{v}_2d\Omega\right],\label{C4}
\end{split}
\end{equation}
We must now evaluate \eqref{C4} when Coulomb collisions result in small deflections, i.e. $\chi<<1$. 
Since the modulus $u$ is preserved during a collision, the relative velocity $\bol{u}'=\bol{u}+\delta\bol{u}$ after the collision satisfies
\begin{equation}
\frac{\bol{u}'\cdot\bol{u}}{u^2}=\cos\chi,~~~~\frac{\abs{\bol{u}'\times\bol{u}}}{u}=u\sin\chi.
\end{equation}
Therefore, denoting with $\bol{n}_1$ and $\bol{n}_2$ unit vectors such that $\lr{\bol{u}/u,\bol{n}_1,\bol{n}_2}$ defines an orthonormal set of basis vectors, we may write
\begin{equation}
\delta\bol{u}=\lr{\cos\chi-1}\bol{u}+u\sin\chi\lr{\cos\phi\,\bol{n}_1+\sin\phi\,\bol{n}_2}.
\end{equation}
Expanding trigonometric functions of the deflection angle $\chi$, at leading order one thus finds
\begin{equation}
\int\frac{\delta\bol{u}\delta\bol{u}}{\sin^4\lr{\chi/2}}\sin\chi d\chi d\phi\approx 16\pi \lr{u^2I-\bol{u}\bol{u}} \log\lr{\frac{\chi_{\rm max}}{\chi_{\rm min}}},
\end{equation}
where $\chi_{\rm max}$ and $\chi_{\rm min}$ denote the maximum and minimum angles of deflection that are allowed, and $I$ denotes the identity matrix.  
In conclusion, the collision operator \eqref{C4} reduces to the Landau collision operator
\begin{equation}
\begin{split}
\mc{C}\lr{f_1,f_2}=&2\pi\lr{\frac{e^2}{4\pi\epsilon_0m}}^2\log\lr{\frac{\chi_{\rm max}}{\chi_{\rm min}}}\frac{\p}{\p\bol{v}_1}\cdot\left[\int f_1f_2u^{-1}\lr{I-\frac{\bol{u}\bol{u}}{u^2}}\cdot\lr{\frac{\p\log f_1}{\p\bol{v}_1}-\frac{\p\log f_2}{\p\bol{v}_2}}
d\bol{v}_2\right].\label{Landau}
\end{split}
\end{equation}
For completeness, we recall that the quantity $\log\Lambda=\log\lr{\frac{\chi_{\rm max}}{\chi_{\rm min}}}$ is the Coulomb logarithm, and observe that the interaction tensor $\Pi$ 
encountered in \eqref{Pi} has been reduced to 
the projector $I-\bol{u}\bol{u}/u^2$ onto the space orthogonal to the relative velocity $\bol{u}$ multiplied by $1/u$.
This is the reason why in this setting the Maxwell-Boltzmann distribution $f\propto\exp\lrc{-\beta\lr{m\bol{v}^2/2+\Phi\lr{\bol{x}}}}$ belongs to the kernel of $\Pi$, and thus provides an equilibrium solution of the evolution equation \eqref{ft}.

In the remaining part of this section we wish to verify the identity \eqref{VInt}, which was used to simplify the expansion of the collision integral, in the case of binary Coulomb collisions with small deflections. 
Recalling equation \eqref{svc}, the integral on the left-hand side of equation \eqref{VInt} can be written as
\begin{equation}
\begin{split}
\int&\left[\mc{V}\delta\bol{z}-\frac{1}{2}\lr{\frac{\p}{\p\bol{z}_1}-\frac{\p}{\p\bol{z}_2}}\cdot\lr{\mc{V}\delta\bol{z}\delta\bol{z}}\right]d\bol{z}_1'd\bol{z}_2'\\=&m{\delta\lr{\bol{q}_1-\bol{q}_2}}\int\left[\sigma u\delta\bol{v}_1-\frac{1}{2}\lr{\frac{\p}{\p\bol{v}_1}-\frac{\p}{\p\bol{v}_2}}\cdot\lr{\sigma u\delta\bol{v}_1\delta\bol{v}_1}\right]d\bol{v}_1'd\bol{v}_2'\\
=&
\frac{m\kappa}{2}\delta\lr{\bol{q}_1-\bol{q}_2}\lrs{\int\frac{\delta\bol{u}}{u^3\sin^4\lr{\chi/2}}\,d\Omega-\frac{1}{4}\lr{\frac{\p}{\p\bol{v}_1}-\frac{\p}{\p\bol{v}_2}}\cdot\int\frac{\delta\bol{u}\delta\bol{u}}{u^3\sin^4\lr{\chi/2}}\,d\Omega}\\
\approx&
\frac{m\kappa}{2}\delta\lr{\bol{q}_1-\bol{q}_2}\lrs{-2\frac{\bol{u}}{u^3}\int\frac{d\Omega}{\sin^2\lr{\chi/2}}-\frac{1}{2}\frac{\p}{\p\bol{u}}\cdot 16\pi u^{-1}\lr{I-\frac{\bol{u}\bol{u}}{u^2}}\log\lr{\frac{\chi_{\rm max}}{\chi_{\rm min}}}}\\=
&8\pi m\kappa\delta\lr{\bol{q}_1-\bol{q}_2}\log\lr{\frac{\chi_{\rm max}}{\chi_{\rm min}}}u^{-3}\lr{-\bol{u}+\bol{u}}=\bol{0},
\end{split}
\end{equation}
where we set $\kappa=\lr{e^2/2\pi\epsilon_0m}^2/4$ and approximated integrals involving the small deflection angle $\chi$ by Taylor expansion around $\chi=0$.

\section{Metriplectic structure}
The aim of this section is to show that the derived collision operator \eqref{C6} for collisions occurring in noncanonical phase spaces and the corresponding evolution equations  \eqref{f1t} and \eqref{f2t} for the distribution functions exhibit a metriplectic structure \cite{Morrison1984,Morrison1986}, i.e. an algebraic bracket formalism joining Hamiltonian and dissipative dynamics in consistency with the first and second laws of thermodynamics. 
This result corroborates the idea that the metriplectic bracket 
encapsulates the fundamental geometric structure of the phase space of mechanical systems that are subject to dissipation, in the same way the Poisson bracket characterizes the phase space of ideal systems. 
To this end, we first review some basic aspects pertaining to the notion of metriplectic bracket (the definitions given below can be found in \cite{Sato2020}). 

Let $\mf{X}$ denote a vector space over $\mathbb{R}$. 
A Poisson bracket is a binary operation $\left\{\cdot,\cdot\right\}_{\ast}:\mf{X}^\ast\cp\mf{X}^\ast\rightarrow\mf{X}^\ast$ on the set $\mf{X}^\ast$ of differentiable functionals $F:\mf{X}\rightarrow\mathbb{R}$ satisfying the following axioms 
\begin{subequations}
\begin{align}
&\left\{aF+bG,H\right\}_{\ast}=a\lrc{F,H}_{\ast}+b\lrc{G,H}_{\ast},~~~~\left\{H,aF+bG\right\}_{\ast}=a\lrc{H,F}_{\ast}+b\lrc{H,G}_{\ast},\\
&\lrc{F,F}_{\ast}=0,\\
&\lrc{F,G}_{\ast}=-\lrc{G,F}_{\ast},\\
&\lrc{FG,H}_{\ast}=F\lrc{G,H}_{\ast}+\lrc{F,H}_{\ast}G,\\
&\lrc{F,\lrc{G,H}_{\ast}}_{\ast}+\lrc{G,\lrc{H,F}_{\ast}}_{\ast}+\lrc{H,\lrc{F,G}_{\ast}}_{\ast}=0,
\end{align}
\end{subequations}\label{PAxioms}
for all $a,b\in\mathbb{R}$ and $F,G,H\in\mf{X}^{\ast}$. 
The asterisk appearing in $\lrc{\cdot,\cdot}_{\ast}$ 
is used to distinguish this Poisson bracket, which acts on 
elements of $\mf{X}^{\ast}$, from the Poisson bracket $\lrc{\cdot,\cdot}:\mf{X}\cp \mf{X}\rightarrow \mf{X}$ acting on  functions $f,g\in \mf{X}$ as in equation \eqref{PB}. For the purpose of the present study, we may take  $\mf{X}=C^{\infty}\lr{\Omega}$. 
The five axioms \eqref{PAxioms} are referred to as bilinearity, alternativity, antisymmetry (which follows from the first two axioms by evaluating $\lrc{F+G,F+G}$), Leibniz rule, and Jacobi identity. The meaning of these axioms is the following: 
bilinearity ensures that the Poisson bracket defines an algebra over $\mathbb{R}$, alternativity physically expresses conservation of energy $\mf{H}$ since $\dot{\mf{H}}=\lrc{\mf{H},\mf{H}}=0$, the Leibniz rule implies
that the Poisson bracket behaves as a differential operator, while the Jacobi identity assigns the phase space structure (the Poisson bracket defines a Lie algebra). 

While the Poisson bracket describes ideal dynamics, dissipative dynamics can be represented with the aid of a dissipative bracket $\lrs{\cdot,\cdot}_{\ast}:\mf{X}^{\ast}\times\mf{X}^{\ast}\rightarrow\mf{X}^{\ast}$, which is a binary operation satisfying the following axioms:
\begin{subequations}
\begin{align}
&\lrs{aF+bG,H}_{\ast}=a\lrs{F,H}_{\ast}+b\lrs{G,H}_{\ast},~~~~\lrs{H,aF+bG}_{\ast}=a\lrs{H,F}_{\ast}+b\lrs{H,G}_{\ast},\\
&\lrs{F,F}_{\ast}\geq 0,\\
&\lrs{F,G}_{\ast}=\lrs{G,F}_{\ast},\\
&\lrs{FG,H}_{\ast}=F\lrs{G,H}_{\ast}+\lrs{F,H}_{\ast}G,
\end{align}
\end{subequations}\label{DAxioms}
for all $a,b\in\mathbb{R}$ and $F,G,H\in\mf{X}^{\ast}$. 
These axioms are bilinearity, non-negativity, symmetry, and Leibniz rule. 
We observe that non-negativity 
is associated with entropy growth because within the metriplectic formalism the evolution of entropy $S$ obeys $\dot{S}=\lrs{S,S}_{\ast}\geq 0$. Furthermore, we will see that the symmetry axiom can be physically interpreted in relation to the symmetry of the microscopic process at the origin of dissipation.  

A metriplectic bracket $\lr{\cdot,\cdot,\cdot}_{\ast}:\mf{X}^{\ast}\times\mf{X}^{\ast}\times\mf{X}^{\ast}\rightarrow\mf{X}^\ast$ 
is a ternary operation combining ideal and dissipative dynamics  according to \cite{Morrison1984}
\begin{equation}
\frac{dF}{dt}=\lr{F,S,\mf{H}}_{\ast}=\lrs{F,S}_{\ast}+\lrc{F,\mf{H}}_{\ast},\label{met}
\end{equation}
where the generating functionals $S,\mf{H}\in\mf{X}^{\ast}$ 
physically represent the entropy and the energy of the system respectively, while $F\in\mf{X}^{\ast}$ is a physical observable. 
In order to ensure the consitency of metriplectic mechanics \eqref{met} with the laws of thermodynamics, the entropy $S$ and the energy $\mf{H}$ must satisfy
\begin{subequations}
\begin{align}
\frac{dS}{dt}=&\lr{S,S,\mf{H}}_{\ast}=\lrs{S,S}_{\ast}+\lrc{S,\mf{H}}_{\ast}\geq 0,\\
\frac{d\mf{H}}{dt}=&\lr{\mf{H},S,\mf{H}}_{\ast}=\lrs{\mf{H},S}_{\ast}=0.
\end{align}
\end{subequations}
A sufficient set of consistency conditions is therefore given by
\begin{equation}
\lrc{S,\mf{H}}_{\ast}=0,~~~~\lrs{S,\mf{H}}_{\ast}=0.\label{cond1}
\end{equation}
The conditions \eqref{cond1} can be identically satsfied by choosing the entropy $S$ to be a Casimir invariant of the Poisson bracket and the energy $\mf{H}$ a Casimir invariant of the dissipative bracket:
\begin{equation}
\lrc{S,\mf{H}}_{\ast}=0~~~~\forall\mf{H}\in\mf{X}^{\ast},~~~~\lrs{S,\mf{H}}_{\ast}=0~~~~\forall S\in\mf{X}^{\ast}.\label{cond2}
\end{equation}
When equation \eqref{cond2} is satisfied, a single generating function (free energy) $\Sigma=S-\beta\mf{H}$, with $\beta$ a normalization parameter physically representing the inverse temperature of the system, can be used to generate dynamics:
\begin{equation}
\frac{dF}{dt}=\lr{F,\Sigma}_{\ast}=-\beta^{-1}\lrc{F,\Sigma}_{\ast}+\lrs{F,\Sigma}_{\ast}.
\label{met2}
\end{equation}
It turns out that equations \eqref{f1t} and \eqref{f2t} can be written in the form \eqref{met2}. In particular, for this system the Poisson bracket is given by 
\begin{equation}
\begin{split}
\lrc{F,G}_{\ast}=&\int f_1\lrc{\frac{\delta F}{\delta f_1},\frac{\delta G}{\delta f_1}}_1d\bol{z}_1+\int f_2\lrc{\frac{\delta F}{\delta f_2},\frac{\delta G}{\delta f_2}}_2d\bol{z}_2\\=&\int f_1{\frac{\p}{\p \bol{z}_1}\lr{\frac{\delta F}{\delta f_1}}\cdot\lrs{\mc{J}_1\cdot\frac{\p}{\p\bol{z}_1}\lr{\frac{\delta G}{\delta f_1}}}}d\bol{z}_1+\int f_2{\frac{\p}{\p \bol{z}_2}\lr{\frac{\delta F}{\delta f_2}}\cdot\lrs{\mc{J}_2\cdot\frac{\p}{\p\bol{z}_2}\lr{\frac{\delta G}{\delta f_2}}}}d\bol{z}_2.
\end{split}
\end{equation}
Indeed, one can verify that
\begin{equation}
\lrc{f_1\lr{\bol{z}_1',t},\mf{H}_{12}}_{\ast}=\int f_1\lrc{\delta\lr{\bol{z}_1'-\bol{z}_1},H_1+\Phi_1}_1d\bol{z}_1.
\end{equation}
Integrating by parts, for any point $\bol{z}_1\in\Omega$ one thus obtains 
\begin{equation}
\lrc{f_1\lr{\bol{z}_1,t},\mf{H}_{12}}_{\ast}=-\lrc{f_1,H_1+\Phi_1}_1.
\end{equation}
Furthermore, the dissipative bracket can be written as
\begin{equation}
\left[F,G\right]_{\ast}=-\int f_1f_2\Pi_{jk}\lrs{\mc{J}^{ij}_1\frac{\p}{\p z_1^i}\lr{\frac{\delta F}{\delta f_1}}-\mc{J}^{ij}_2\frac{\p}{\p z_2^i}\lr{\frac{\delta F}{\delta f_2}}}\lrs{\mc{J}^{km}_1\frac{\p}{\p z_1^m}\lr{\frac{\delta G}{\delta f_1}}-\mc{J}_2^{km}\frac{\p}{\p z_2^m}\lr{\frac{\delta G}{\delta f_2}}}d\bol{z}_1d\bol{z}_2.\label{DB117}
\end{equation}
Indeed, using this expression gives 
\begin{equation}
\begin{split}
\left[f_1\lr{\bol{z}_1',t},S_{12}\right]_{\ast}=&
-\int f_1f_2\Pi_{jk}\lrs{\mc{J}^{ij}_1\frac{\p}{\p z_1^i}{\delta\lr{\bol{z}_1'-\bol{z}_1}}}\lr{\mc{J}_2^{km}\frac{\p\log f_2}{\p z_2^m}-\mc{J}^{km}_1\frac{\p\log f_1}{\p z_1^m}}d\bol{z}_1d\bol{z}_2.
\end{split}
\end{equation}
Integrating by parts, for any point $\bol{z}_1\in\Omega$ one arrives at 
\begin{equation}
\begin{split}
\left[f_1\lr{\bol{z}_1,t},S_{12}\right]_{\ast}=&
\frac{\p}{\p\bol{z}_1}\cdot\lrs{f_1\mc{J}_1\cdot\int f_2\Pi\cdot\lr{\mc{J}_2\cdot\frac{\p\log f_2}{\p\bol{z}_2}-\mc{J}_1\cdot\frac{\p\log f_1}{\p\bol{z}_1}}d\bol{z}_2}=\mc{C}\lr{f_1,f_2}\lr{\bol{z}_1,t}.
\end{split}
\end{equation}
We have therefore shown that equations \eqref{f1t} and \eqref{f2t} have the metriplectic structure \cite{Morrison1984} below  
\begin{subequations}
\begin{align}
&\frac{\p f_1}{\p t}=\lr{f_1,S_{12},\mf{H}_{12}}_{\ast}=\lrc{f_1,\mf{H}_{12}}_{\ast}+\lrs{f_1,S_{12}}_{\ast},\\
&\frac{\p f_2}{\p t}=\lr{f_2,S_{12},\mf{H}_{12}}_{\ast}=\lrc{f_2,\mf{H}_{12}}_{\ast}+\lrs{f_2,S_{12}}_{\ast}.
\end{align}
\end{subequations}
Furthermore, since $S_{12}$ and $\mf{H}_{12}$ are Casimir invariants of the Poisson bracket and the dissipative bracket respectively, these equations can be cast in the form \eqref{met2} with generating function $\Sigma=S_{12}-\beta\mf{H}_{12}$.  
 


\subsection{Relationship with the curvature-like framework of metriplectic 4-bracket dynamics}
Recently, a 4-bracket formalism 
describing dissipative dynamics 
through curvature-like tensors 
has been proposed 
\cite{PJM23}. 
This framework is inclusive,
i.e. it includes the binary dissipative  bracket discussed above, and also elucidates certain geometric aspects of
dissipation. 

In finite dimensions, the 4-bracket acting on smooth functions $f,g,h,\ell\in C^{\infty}\lr{\Omega}$ is defined via a contravariant 4-tensor with components $R^{ijkl}$ according to
\begin{equation}
\lr{f,g;h,\ell}=R^{ijkl}\frac{\p f}{\p z^i}\frac{\p g}{\p z^j}\frac{\p h}{\p z^k}\frac{\p\ell}{\p z^l}.\label{4M}
\end{equation}
Here, the 4-tensor $R^{ijkl}$ is also assumed to satisfy the symmetries 
\begin{subequations}
\begin{align}
R^{ijkl}=&-R^{jikl},\\
R^{ijkl}=&-R^{ijlk},\\
R^{ijkl}=&R^{klij}.
\end{align}\label{MM}
\end{subequations}
A 4-bracket satisfying  the properties above 
and the positive semi-definiteness condition $\lr{f,g;f,g}\geq 0$ for all $f,g\in C^{\infty}\lr{\Omega}$  
is called minimal metriplectic.
Using \eqref{4M} and \eqref{MM}, one can verify that the 4-bracket is linear in its arguments, that it acts as a derivation on its arguments, and that
\begin{subequations}
\begin{align}
\lr{f,g;h,\ell}&=-\lr{g,f;h,\ell},\\
\lr{f,g;h,\ell}&=-\lr{f,g;\ell,h},\\
\lr{f,g;h,\ell}&=\lr{h,\ell;f,g}.
\end{align}
\end{subequations}
If $R^{ijkl}$ were the fully contravariant form of a 
Riemannian curvature tensor, the first and second Bianchi identities
\begin{subequations}
\begin{align}
R^{ijkl}+R^{iklj}+R^{iljk}&=0,\label{Bianchi1}\\
R^{ijkl}_{;m}+R^{ijlm}_{;k}+R^{ijmk}_{;l}&=0,\label{Bianchi2}
\end{align}\label{4M2}
\end{subequations}
would provide additional structure to the 4-bracket. In particular, the first Bianchi identity implies that
\begin{equation}
\lr{f,g;h,\ell}+\lr{f,h;\ell,g}+\lr{f,\ell;g,h}=0,
\end{equation}
while the second Bianchi identity, whose precise role in the 4-bracket  theory has yet to be clarified,  
offers an analogy with 
the Jacobi identity obeyed by Poisson tensors in Hamiltonian dynamics. 

The dissipative bracket \eqref{DB117} 
can be written as a 4-bracket $\lr{\cdot,\cdot;\cdot,\cdot}_{\ast}:\mf{X}^{\ast}\times\mf{X}^{\ast}\times\mf{X}^{\ast}\times\mf{X}^{\ast}\rightarrow\mf{X}^{\ast}$ on $\mf{X}^{\ast}$. 
To see this, define the 4-bracket acting on $F,G,H,L\in\mf{X}^{\ast}$
\begin{equation}
\lr{F,G;H,L}_{\ast}=\frac{\tau_c^2}{2}\int f_1f_2\Gamma {\mc{J}^{ij}_1\mc{J}^{kl}_1}P\lrs{F_f}_iP\lrs{G_f}_jP\lrs{H_f}_kP\lrs{L_f}_l\,d\bol{z}_1 d\bol{z}_2,\label{4B}
\end{equation}
where
\begin{equation}
P\lrs{F_f}_i=\frac{\p}{\p z_1^i}\frac{\delta F}{\delta f_1}-\frac{\p}{\p z_2^i}\frac{\delta F}{\delta f_2}.
\end{equation}
Recalling that by hypothesis on the spatial 
scale of the interaction $\mc{J}_{1}\approx\mc{J}_2$,  
and using the symmetry of the interaction potential energy \eqref{sym} and the expression of the interaction tensor \eqref{Pi}, we find 
\begin{equation}
\lr{F,\Phi_1;G,\Phi_2}_{\ast}=\lrs{F,G}_{\ast}.
\end{equation}
We observe that the 4-tensor $R^{ijkl}$ in this setting is given by
\begin{equation}
R^{ijkl}={\mc{J}_1^{ij}\mc{J}_{1}^{kl}},\label{RMM}
\end{equation}
which satisfies the  symmetries \eqref{MM}. 
It is worth mentioning that the possibility of decomposing general algebraic curvature tensors in terms of symmetric and asymmetric $2$-tensors is discussed in \cite{Fiedler}. 
We also have
\begin{equation}
R^{ijkl}_{;k}+R^{jlki}_{;k}+R^{likj}_{;k}=-\mc{J}^{lk}_1\frac{\p\mc{J}^{ij}_{1}}{\p z^k}-\mc{J}^{ik}_1\frac{\p\mc{J}^{jl}_{1}}{\p z^k}-\mc{J}^{jk}_1\frac{\p\mc{J}^{li}_{1}}{\p z^k}+\mc{J}_1^{ij}\frac{\p\mc{J}_1^{kl}}{\p z^k}+\mc{J}_1^{jl}\frac{\p\mc{J}_1^{ki}}{\p z^k}+\mc{J}_1^{li}\frac{\p\mc{J}_1^{kj}}{\p z^k}.\label{JIR}
\end{equation}
Therefore, if the coordinates $\bol{z}=\lr{z^1,...,z^n}$ 
are chosen to span the invariant measure assigned by $\mc{J}_1$, the identities \eqref{IM2} hold, and equation \eqref{JIR} can be used to express the Jacobi identity,
\begin{equation}
R^{ijkl}_{;k}+R^{jlki}_{;k}+R^{likj}_{;k}=-\mc{J}^{lk}_1\frac{\p\mc{J}^{ij}_{1}}{\p z^k}-\mc{J}^{ik}_1\frac{\p\mc{J}^{jl}_{1}}{\p z^k}-\mc{J}^{jk}_1\frac{\p\mc{J}^{li}_{1}}{\p z^k}=0.\label{JIR2}
\end{equation}
The definition of the 4-bracket \eqref{4B} can be further ameliorated so that 
the first Bianchi identity \eqref{Bianchi1} holds. 
Indeed, suppose that $\sigma^{ij}$ and $\mu^{kl}$
are two antisymmetric contravariant 2-tensors. Then, one can verify that the contravariant 4-tensor
\begin{equation}
\mc{R}^{ijkl}=\sigma^{ij}\mu^{kl}+\sigma^{il}\mu^{kj}-\sigma^{ki}\mu^{jl}-\sigma^{ji}\mu^{kl},\label{R20}
\end{equation}
satisfies the first Bianchi idenity \eqref{Bianchi1}. This suggests to set $\sigma^{ij}=\mu^{ij}=\mc{J}^{ij}_1$ and to  define the 4-bracket 
\begin{equation}
\lr{F,G;H,L}_{\ast}'=\frac{\tau_c^2}{6}\int f_1f_2\Gamma {\mc{R}^{ijkl}}P\lrs{F_f}_iP\lrs{G_f}_jP\lrs{H_f}_kP\lrs{L_f}_l\,d\bol{z}_1 d\bol{z}_2,\label{4B2}
\end{equation}
with 4-tensor
\begin{equation}
{\mc{R}^{ijkl}}={\mc{J}^{ij}_1\mc{J}^{kl}_1+\mc{J}^{il}_1\mc{J}^{kj}_1-\mc{J}^{ki}_1\mc{J}^{jl}_1-\mc{J}^{ji}_1\mc{J}^{kl}_1}.\label{R2}
\end{equation}
Note that the 4-tensor \eqref{R2} satisfies both the symmetries \eqref{MM} and the first Bianchi identity \eqref{Bianchi1}. 
The form \eqref{R2} is also consistent with the decomposition of algebraic curvature tensors discussed in \cite{Fiedler}. 
Furthermore, the bracket \eqref{4B2} 
correctly reproduces the dissipative bracket associated with the derived  collision operator \eqref{DB117}, 
\begin{equation}
\lr{F,\Phi_1;G,\Phi_2}_{\ast}'=\lrs{F,G}_{\ast}.
\end{equation}
We remark that however the second Bianchi identity \eqref{Bianchi2} is not satisfied by \eqref{R2}. 
Nevertheless, 
one can verify that 
\begin{equation}
\mc{R}^{ijkl}_{;k}+\mc{R}^{jlki}_{;k}+\mc{R}^{likj}_{;k}=\lr{3-3}\lr{\mc{J}^{lk}_1\frac{\p\mc{J}^{ij}_{1}}{\p z^k}+\mc{J}^{ik}_1\frac{\p\mc{J}^{jl}_{1}}{\p z^k}+\mc{J}^{jk}_1\frac{\p\mc{J}^{li}_{1}}{\p z^k}}=0.\label{JIR3}
\end{equation}
Note that now equation \eqref{JIR3} holds for any antisymmetric tensor $\mc{J}_1$, regardless of whether $\mc{J}_1$ satisfies the Jacobi identity. 

These results suggest that an alternative set of axioms for a general contravariant 4-tensor $R^{ijkl}$ generating metriplectic 4-bracket dynamics is given by
\begin{subequations}
\begin{align}
R^{ijkl}&=-R^{jikl},\\
R^{ijkl}&=-R^{ijlk},\\
R^{ijkl}&=R^{klij},\\
R^{ijkl}+R^{iklj}+R^{iljk}&=0,\\
{R}^{ijkl}_{;k}+{R}^{jlki}_{;k}+{R}^{likj}_{;k}&=0,\label{GJI}
\end{align}
\end{subequations}
where the second Bianchi identity \eqref{Bianchi2} has been replaced by the `extended Jacobi identity' \eqref{GJI}. It is worth observing that equation \eqref{GJI} can be equivalently written as
\begin{equation}
\frac{\p}{\p z^k}\lr{f,g;z^k,h}+\circlearrowright=\lr{\frac{\p f}{\p z^k},g;z^k,h}+\lr{f,\frac{\p g}{\p z^k};z^k,h}+\lr{f,g;\frac{\p z^k}{\p z^k},h}+\lr{f,g;z^k,\frac{\p h}{\p z^k}}+\circlearrowright,
\end{equation}
where $\circlearrowright$ denotes summation of even permutations of $f$, $g$, and $h$. 
This expression is similar to the distributive property of time derivatives following from the fundamental identity proposed as replacement for the Jacobi identity in Nambu mechanics \cite{Tak,SatoPTEP}.
We conclude by observing that the 
4-tensor \eqref{R20} is reminiscent of the Kulkarni-Nomizu construction 
for symmetric contravariant 2-tensors $\alpha^{ij}$ and $\beta^{kl}$ where
a 4-tensor with the algebraic symmetries \eqref{MM} and \eqref{4M2} of the Riemannian curvature tensor is defined through the Kulkarni-Nomizu product as follows:
\begin{equation}
\mc{R}^{ijkl}=\alpha^{ik}\beta^{jl}-\alpha^{il}\beta^{jk}+\beta^{ik}\alpha^{jl}-\beta^{il}\alpha^{jk}.
\end{equation}


\section{Concluding remarks}

In this paper, we have derived a collision operator \eqref{C7} for fast and spatially localized interactions (with respect to unperturbed dynamics) in noncanonical phase space.
The interaction force driving scattering events satisfies the elastic scattering condition \eqref{dEdt3} and the symmetry condition \eqref{sym}, but it is otherwise of arbitrary nature. 
The derived collision operator is consistent with conservation of particle number and energy, it satisfies an H-theorem, and it preserves the interior  Casimir invariants induced by the microscopic Poisson tensor 
on the field theory. Furthermore, it reduces to the Landau collision operator in the limit of small deflection binary Coulomb collisions in canonical phase space, and it provides the evolution equation for the distribution function with a metriplectic structure. This result supports the idea that the metriplectic bracket captures basic properties associated with the algebraic structure of dissipative systems.

The shape of thermodynamic equilibria \eqref{MaxEnt4} resulting from maximization of entropy departs  from Maxwell-Boltzmann statistics: the distribution function depends on 
the Casimir invariants associated with the noncanonical phase space structure, and 
the Jacobian connecting the preserved phase space measure with the configuration space measure.  

As discussed in section 5, the present theory can be applied to explain self-organizing phenomena driven by the noncanonical Hamiltonian structure of the phase space, as well as to describe collisionless relaxation in magnetized plasmas and stellar systems by modeling interactions in terms of collisions between clusters of charged particles or gravitating bodies. 
In particular, conservation of Casimir and/or macroscopic invariants during the equilibration of the statistical ensemble explains, in consistency with the laws of thermodynamics, the self-organization of stable structures in reduced mechanical systems affected by rigidity constraints, as well as the creation of thermal equilibria in collisionless systems over time scales shorter than the binary Coulomb or gravitational collision times between charged particles or massive bodies.

\section*{Acknowledgment}
NS is grateful to M. Yamada for useful discussion. 

\section*{Statements and declarations}

\subsection*{Data availability}
Data sharing not applicable to this article as no datasets were generated or analysed during the current study.

\subsection*{Funding}
The research of NS was partially supported by JSPS KAKENHI Grant No. 21K13851 and 22H04936.
PJM was supported by U.S. Dept. of Energy Contract \# DE-FG05-80ET-53088  and a Humboldt Foundation
Research Award.

\subsection*{Competing interests} 
The authors have no competing interests to declare that are relevant to the content of this article.




\begin{thebibliography}{99}

\bibitem{Morrison98} P. J. Morrison, 
\ti{Hamiltonian description of the ideal fluid}, 
Rev. Mod. Phys. \tb{70}, pp. 467-521 (1998). 

\bibitem{Morrison82} P. J. Morrison,
\ti{Poisson brackets for fluids and plasmas}, in Mathematical Methods in Hydrodynamics and Integrability in Dynamical Systems, Tabor M., Treve Y. (Eds.), American Institute of Physics Conference Proceedings, No. 88, American Institute of Physics, New York, pp. 13-46 (1982).

\bibitem{YosPRL} Z. Yoshida, H. Saitoh, J. Morikawa, 
Y. Yano, S. Watanabe, and Y. Ogawa, 
\ti{Magnetospheric Vortex Formation: Self-Organized Confienement of Charged Particles},
Phys. Rev. Lett. \tb{104}, 235004 (2010). 

\bibitem{Kawazura19} Y. Kawazura, M. Barnes, and A. A. Schekochihin, 
\ti{Thermal disequilibration of ions and electrons by collisionless plasma turbulence}, 
PNAS \tb{116}, 3, pp. 771-776 (2019).

\bibitem{Lynden} D. Lynden-Bell, 
\ti{Statistical mechanics of violent relaxation in stellar systems}, 
Mon. Not. R. Astr. Soc. \tb{136}, pp. 101-121 (1967).

\bibitem{Yos2014} Z. Yoshida, S. M. Mahajan, 
\ti{Self-organization in foliated phase space: Construction of a scale hierarchy by adiabatic invariants of magnetized particles}, 
Prog. Theor. Exp. Phys. \tb{2014}, 073J01 (2014).

\bibitem{Chavanis22} P.-H. Chavanis, 
\ti{Kinetic theory of collisionless relaxation for systems with long-range interactions}, 
Physica A, 128089 (2022).

\bibitem{Kadomstev} B. B. Kadomstev and O. P. Pogutse,
\ti{Collisionless relaxation in systems with Coulomb interactions}, Phys. Rev. Lett. \tb{25}, 17 (1970). 

\bibitem{MorrisonMV} 
P. J. Morrison, 
\ti{The Maxwell-Vlasov Equations as a Continuous Hamiltonian System},
Phys. Lett. \tb{80A}, 5-6 (1980).

\bibitem{Marsden82} J. E. Marsden, A. Weinstein, 
\ti{The Hamiltonian structure of the Maxwell-Vlasov equations}, Physica D: Nonlinear Phenomena \tb{4}, pp. 394-406 (1982).

\bibitem{Landau1936} 
L.D. Landau, \ti{Die kinetische Gleichung für den Fall Coulombscher Wechselwirkung}, 
Phys. Z. Sowjetunion \tb{10}, 2, pp. 154-164 (1936).

\bibitem{Kampen} N. G. Van Kampen and B. U. Felderhof, \ti{Collisions}, in 
Theoretical Methods in Plasma Physics, North-Holland, Amsterdam, pp. 188-208 (1967). 

\bibitem{Lenard} A. Lenard, 
\ti{On Bogoliubov’s kinetic equation for a spatially homogeneous plasma}, Ann. Phys. \tb{10},
390 (1960).

\bibitem{Chavanis04} P.-H. Chavanis, 
\ti{Generalized thermodynamics and kinetic equations: Boltzmann, Landau, Kramers, 
and Smoluchowski}, Physica A 332, pp. 89-122 (2004).

\bibitem{Presse} S. Press\'e, K. Ghosh, J. Lee, and K. A. Dill,
\ti{Nonadditive entropies yield probability distributions with biases not warranted by the data},
Phys. Rev. Lett. \tb{111}, 180604 (2013).

\bibitem{Sudar} E. Sudarshan and N. Mukunda, \ti{Classical Dynamics: A
Modern Perspective}, Wiley, New York (1974).

\bibitem{Littlejohn} R. Littlejohn, \ti{Singular Poisson tensors}, in Mathematical Methods in Hydrodynamics and Integrability in Dynamical Systems, Tabor M., Treve Y. (Eds.), American Institute of Physics Conference Proceedings, No. 88, American Institute of Physics, New York, pp. 47-66 (1982). 

\bibitem{Moore} C. C. Moore, 
\ti{Ergodic theorem, ergodic theory, and statistical mechanics}, 
Proc. Natl. Acad. Sci. \tb{112}, pp. 1907-1911 (2015).

\bibitem{Shannon} C. E. Shannon, 
\ti{A mathematical theory of communication},  Bell Syst. Techn. J. \tb{27}, pp. 379-423 (1948).

\bibitem{Jaynes} E. T. Jaynes, 
\ti{Information theory and statistical mechanics}, 
Phys. Rev. \tb{106}, pp. 620-630 (1957).

\bibitem{Sato2016} N. Sato and Z. Yoshida, 
\ti{Up-hill diffusion, creation of density gradients: Entropy measure for systems with topological constraints}, Phys. Rev. E \tb{93}, 062140 (2016). 

\bibitem{Chavanis06} P.-H. Chavanis, 
\ti{Quasi-stationary states and incomplete violent relaxation in systems with
long-range interactions}, Physica A \tb{365}, pp. 102-107 (2006).

\bibitem{Jung} S. Jung, P. J. Morrison, and H. L. Swinney,
\ti{Statistical mechanics of two-dimensional turbulence}, 
J. Fluid Mech. \tb{554}, 
pp. 433-456 (2006). 

\bibitem{Chavanis96} P.-H. Chavanis, J. Sommeria, and R. Robert, \ti{Statistical mechanics of two-dimensional vortices and collisionless stellar systems}, Astrophys. J. \tb{471}, pp. 385-399 (1996).

\bibitem{Marsden} J. E. Marsden, P. J. Morrison, and A. Weinstein, 
\ti{The Hamiltonian structure of the BBGKY hierarchy equations}, 
Contemporary Mathematics \tb{28}, pp. 115-124 (1984).

\bibitem{Ewart} R. J. Ewart, A. Brown, T. Adkins, and A. A. Schekochihin, 
\ti{Collisionless relaxation of a Lynden-Bell plasma}, 
J. Plasma Phys. \tb{88}, 925880501 (2022).

\bibitem{Vallis} G. K. Vallis, G. F. Carnevale, W. R. Young, 
\ti{Extremal energy properties and construction of stable solutions of the Euler equations}, 
J. Fluid Mech. \tb{207}, pp. 133-152 (1989).

\bibitem{Bloch} A. M. Bloch, R. W. Brocket, T. S. Ratiu, 
\ti{Completely integrable gradient flows}, 
Comm. Math. Phys. \tb{147}, pp. 57-74 (1992).

\bibitem{Fl} 
G. R. Flierl and P. J. Morrison, \ti{Hamiltonian-Dirac Simulated Annealing: Application to the Calculation of Vortex States},  Physica D \t{240}, pp. 212-232 (2011).

\bibitem{Furukawa} M. Furukawa, T. Watanabe, P. J. Morrison, K. Ichiguchi, 
\ti{Calculation of large-aspect-ratio tokamak and toroidally-averaged stellarator equilibria of high-beta reduced magnetohydrodynamics via simulated annealing}, Phys. Plasmas \tb{25}, 082506 (2018). 






\bibitem{Morrison1984} P. J. Morrison,
\ti{Bracket formulation for irreversible classical fields}, 
Physics Letters A \tb{100}, pp. 423-427 (1984). 

\bibitem{Morrison1986} P. J. Morrison, \ti{A paradigm for joined Hamiltonian and dissipative systems}, 
Physica D \tb{18}, 1-3, pp. 410-419 (1986). 


\bibitem{Materassi} M. Materassi, E. Tassi, 
\ti{Metriplectic framework for dissipative magneto-hydrodynamics}, Physica D: Nonlinear Phenomena \tb{241}, pp. 729-773 (2012).

\bibitem{Coquinot} B. Coquinot, P. J. Morrison, 
\ti{A general metriplectic framework with application to dissipative extended magnetohydrodynamics}, J. Plasma Physics \tb{86}, 3 (2020).

\bibitem{Sato2020} N. Sato, 
\ti{Dissipative brackets for the Fokker-Planck equation in Hamiltonian systems and characterization of metriplectic manifolds}, 
Physica D \tb{411}, 132571 (2020). 

\bibitem{Sche} A. A. Schekochihin, S. C. Cowley, 
W. Dorland, G. W. Hammett, G. G. Howes, E. Quataert, T. Tatsuno, \ti{Astrophysical gyrokinetics: kinetic and fluid turbulent cascades in
magnetized weakly collisional plasmas}, Astrophys. J. Suppl. Ser. \tb{182}, pp. 310-377 (2009).

\bibitem{Kawazura20} Y. Kawazura, A. A. Schekochihin, M. Barnes, J. M. TenBarge, Y. Tong, K. G. Klein, and W. Dorland, 
\ti{Ion versus electron heating in compressively driven astrophysical gyrokinetic turbulence}, 
Phys. Rev. X \tb{10}, 041050 (2020). 

\bibitem{Hirv} E. Hirvijoki, A. J. Brizard, and David Pfefferl\'e, 
\ti{Differential formulation of the gyrokinetic
Landau operator},
J. Plasma Phys. \tb{83}, 595830102 (2017).

\bibitem{Burby} J. W. Burby, A. J. Brizard, and H. Qin,
\ti{Energetically consistent collisional gyrokinetics},
Phys. Plasmas \tb{22}, 100707 (2015).



\bibitem{Darboux1} V. I. Arnold, in Mathematical Methods of Classical Mechanics, 2nd ed., Springer, New York, pp. 230-232 (1989).

\bibitem{Darboux2} M. de Le\'on, 
in Methods of Differential Geometry in Analytical Mechanics, 
Elsevier, New York, pp. 250-253 (1989). 

\bibitem{PJMHopfI} P. J. Morrison and G. I. Hagstrom, \ti{Continuum Hamiltonian Hopf Bifurcation I},  in Nonlinear Physical Systems – Spectral Analysis, Stability and Bifurcations, eds. O. Kirillov and D. Pelinovsky, Wiley (2014).


\bibitem{Hasegawa} A. Hasegawa and K. Mima, \ti{A Pseudo-Three-Dimensional Turbulence in Magnetized Nonuniform Plasma}, Phys. Fluids \tb{21}, 1, pp. 87-92 (1977).

\bibitem{NSAIP} N. Sato and Z. Yoshida, 
\ti{Charged particle diffusion in a magnetic dipole trap}, AIP Conf. Proc. \tb{1928}, 020014 (2018).

\bibitem{SatoPRE2} N. Sato and Z. Yoshida, 
\ti{Diffusion with finite-helicity field tensor: A mechanism of generating heterogeneity}, Phys. Rev. E \tb{97}, 2, 022145 (2018).   

\bibitem{Chandre} C. E. Caligan and C. Chandre, 
\ti{Conservative dissipation: How important is the Jacobi identity in the dynamics}, 
Chaos \tb{26}, 053101 (2016). 


\bibitem{Boxer} A. C. Boxer, R. Bergmann, J. L. Ellsworth, 
D. T. Garnier, J. Kesner, M. E. Mauel, and P. Woskov,
\ti{Turbulent inward pinch of plasma confined by a levitated dipole magnet}, 
Nature Physics \tb{6}, pp. 207-212 (2010). 

\bibitem{Ken} 
N. Kenmochi, Y. Yokota, M. Nishiura, H. Saitoh, N. Sato, K. Nakamura, T. Mori, K. Ueda, and Z. Yoshida, 
\ti{Inward diffusion driven by low frequency fluctuations in self-organizing magnetospheric plasma}, 
Nucl. Fusion \tb{62}, 026041 (2022).


\bibitem{Helander2014} P. Helander, 
\ti{Microstability of magnetically confined electron-positron plasmas}, 
Phys. Rev. Lett. \tb{113}, 135003 (2014). 

\bibitem{Sato2023} N. Sato,
\ti{Maximum Entropy States of Collisionless Positron-Electron Plasma in a Dipole Magnetic Field}, Physics of Plasmas \tb{30}, 042503 (2023).

\bibitem{Goldston} 
R. J. Goldston and P. H. Rutherford  
in \ti{Introduction to Plasma Physics} (IOP)
p. 173 (1995).

\bibitem{Cary} J. R. Cary and A. J. Brizard, 
\ti{Hamiltonian theory of guiding-center motion}, 
Rev. Mod. Phys. \tb{81}, pp. 693-738 (2009).

\bibitem{LandauMec} L. D. Landau and E. M. Lifshitz  
in Mechanics, Butterworth-Heinemann, Oxford, 3rd ed., pp. 154-167 (1976).

\bibitem{Maoz} D. Maoz, 
in Astrophysics in a nutshell, Princeton University Press, 2nd ed., 
pp. 189-190 (2016).

\bibitem{PJM23} P. J. Morrison and M. H. Updike, 
\ti{An inclusive curvature-like framowork describing dissipation: metriplectic 4-bracket dynamics}, 	arXiv:2306.06787.

\bibitem{Fiedler} B. Fiedler, 
\ti{
Determination of the structure of algebraic curvature tensors by means of Young symmetrizers}, 
Séminaire Lotharingien de Combinatoire \tb{48}, B48d (2002).

\bibitem{Tak} L. Takhtajan, 
\ti{On foundation of the generalized Nambu mechanics}, Comm. Math. Phys. \tb{160}, 
pp. 295-315 (1994).
 
\bibitem{SatoPTEP} N. Sato
\ti{Generalization of Hamiltonian mechanics to a three-dimensional phase space}, 
Prog. Theor. Exp. Phys. \tb{2021}, 6, 063A01 (2021). 


\bibitem{Dupree72} 
T. H. Dupree,
\ti{Theory of Phase Space Density Granulation in Plasma},
Phys. Fluids \tb{15}, 
pp. 15, 334–344 (1972).

\bibitem{Rostoker} N. Rostoker, 
\ti{Superposition of Dressed Test Particles}, 
Phys. Fluids \tb{7}, pp. 479-490 (1964). 



































\end{thebibliography}
\end{document}